\documentclass[prx, reprint, superscriptaddress]{revtex4-2}
\usepackage{amsmath}
\usepackage{amssymb}
\usepackage{graphicx}
\usepackage[caption=false, position=top, singlelinecheck=off, justification=raggedright]{subfig}
\usepackage{multirow}
\usepackage{color}
\usepackage{pst-node}
\usepackage[unicode]{hyperref}
\hypersetup{
	unicode=true,         	
	colorlinks=true,      	
	linkcolor=blue,		   	
	citecolor=blue,       	
	urlcolor=blue		   	
}

\begin{document}

\title{Higgs time crystal in a high-$T_c$ superconductor}

\author{Guido Homann}
\email{ghomann@physnet.uni-hamburg.de}
\affiliation{Zentrum f\"ur Optische Quantentechnologien and Institut f\"ur Laserphysik, 
Universit\"at Hamburg, 22761 Hamburg, Germany}

\author{Jayson G. Cosme}
\affiliation{Zentrum f\"ur Optische Quantentechnologien and Institut f\"ur Laserphysik, 
Universit\"at Hamburg, 22761 Hamburg, Germany}
\affiliation{The Hamburg Centre for Ultrafast Imaging, Luruper Chaussee 149, 22761 Hamburg, Germany}
\affiliation{National Institute of Physics, University of the Philippines, Diliman, Quezon City 1101, Philippines}

\author{Ludwig Mathey}
\affiliation{Zentrum f\"ur Optische Quantentechnologien and Institut f\"ur Laserphysik, 
Universit\"at Hamburg, 22761 Hamburg, Germany}
\affiliation{The Hamburg Centre for Ultrafast Imaging, Luruper Chaussee 149, 22761 Hamburg, Germany}

\date{\today}
\begin{abstract}
We propose to induce a time crystalline state in a high-$T_c$ superconductor, by optically driving a sum resonance of the Higgs mode and a Josephson plasma mode. The generic cubic process that couples these fundamental excitations converts driving of the sum resonance into simultaneous resonant driving of both modes, resulting in an incommensurate subharmonic motion. We use a numerical implementation of a semi-classical driven-dissipative lattice gauge theory on a three-dimensional layered lattice, which models the geometry of cuprate superconductors, to demonstrate the robustness of this motion against thermal fluctuations. We demonstrate this light-induced time crystalline phase for mono- and bilayer systems and show that this order can be detected for pulsed driving under realistic technological conditions.
\end{abstract}
\maketitle

\section{Introduction}
Optical driving of solids constitutes a new method of designing many-body states. Striking examples of this approach include light-induced superconductivity \cite{Fausti2011, Hu2014, Cremin2019} as well as optical control of charge density wave phases \cite{Kogar2019}. For these states, the carefully tuned light field either renormalises the phase boundary of the equilibrium phase, as is the case for light-induced superconductivity, or renormalises a near-by metastable state into a stable state of the driven system, as is the case for light-controlled charge density waves.

These observations are part of a larger effort to determine the steady states of periodically driven many-body systems. In a parallel development in cold atom systems, serving as well-defined many-body toy models, the generic regimes that were proposed, see Refs.~\cite{Cosme2018,Georges2018}, firstly include renormalised equilibrium states, for which the above mentioned states are examples. Secondly, regimes beyond the equilibrium states emerge, in particular genuine non-equilibrium orders, which have no equilibrium counterpart, and only exist in the driven state. A striking example of a non-equilibrium order is time crystals \cite{Wilczek2013,Sacha2018,Gambetta2018,Buca2019,Heugel2019,Else2020,Yao2020}, reported in systems such as ion traps or nitrogen-vacancy centers \cite{Choi2017,Zhang2017}. 
Thirdly, for strong driving, chaotic states emerge. These different regimes are achieved for different driving amplitudes and driving frequencies, which constitutes the dynamical phase diagram of the system.

In this paper, we propose to create a light-induced time crystalline state in a high-$T_c$ superconductor. This advances light control of superconductors towards genuine non-equilibrium orders, and furthers time crystals in the solid state domain \cite{Chew2020}. We characterise the observed non-equilibrium state as a time crystal based on the following criteria \cite{Else2020}: (i) A time crystal spontaneously breaks time-translation symmetry, that is, it exhibits a subharmonic response to the drive. (ii) The subharmonic response is robust against perturbations which respect the time-translation symmetry of the Hamiltonian. (iii) The subharmonic response emerges in a many-body system with a large number of locally coupled degrees of freedom, and it persists for an infinite time.

\begin{figure}[!t]
	\centering
	\includegraphics[scale=1]{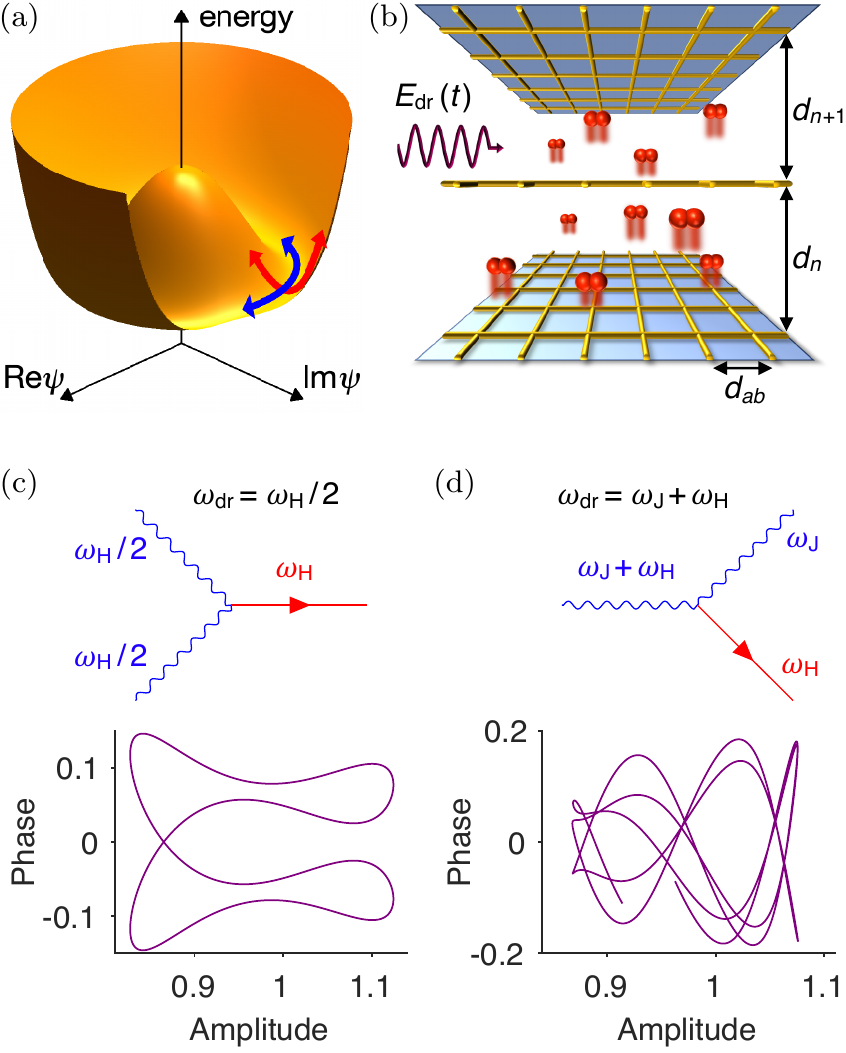}
	\caption{Exciting the Higgs and plasma modes. (a) Illustration of the free energy of a state with broken $U(1)$ symmetry. (b) Illustration of a driven cuprate superconductor modelled as a $U(1)$ gauge theory on an anisotropic lattice. In-plane dynamics is captured by discretising the condensate field in the $ab$-plane. (c) The Higgs mode can be excited resonantly with a driving frequency of $\omega_{\mathrm{dr}} \lesssim \omega_{\mathrm{H}}/2$, by utilising the non-linear coupling between the electromagnetic field and the Higgs field. Upper panel: Diagrammatic representation of the non-linear process. Lower panel: Exemplary dynamical portrait of the phase difference between the superconducting layers and the condensate amplitude in the steady state for 10 driving cycles at zero temperature. 
	(d) We propose to utilise the same non-linear coupling to induce a time-crystalline state by driving the sum resonance of the system at $\omega_{\mathrm{dr}}= \omega_{\mathrm{H}}+\omega_{\mathrm{J}}$. The phase-space trajectories shown in the lower panels of (c) and (d) are obtained using a Josephson junction model for a monolayer cuprate with Higgs frequency $\omega_{\mathrm{H}}/2\pi \approx6.3$ THz and plasma frequency $\omega_{\mathrm{J}}/2\pi \approx16.0$ THz, see Table~\ref{tab:parameters} for full parameter set.}
	\label{fig:1} 
\end{figure}

We call the novel dynamical phase a Higgs time crystal because we induce it via optical driving of a sum resonance of the Higgs mode and a Josephson plasma mode. The Higgs mode and the Josephson plasma mode correspond to the two fundamental collective excitations of a system with broken $U(1)$ symmetry and with an underlying approximate particle-hole symmetry. The Higgs mode is an amplitude oscillation of the order parameter, as depicted in Fig.~\ref{fig:1}(a) for the $|\psi|^4$ theory used in the following. The Higgs mode is a gapped excitation due to the increase of the potential energy in the radial direction. The Josephson plasma mode is a phase oscillation, as indicated. This mode also has a gapped excitation spectrum owing to the electromagnetic interaction of the system. Because of the approximate particle-hole symmetry, these two oscillations are orthogonal to each other \cite{Varma2002, Pekker2015}.

\begin{figure*}[!t]
	\centering
	\includegraphics[scale=1]{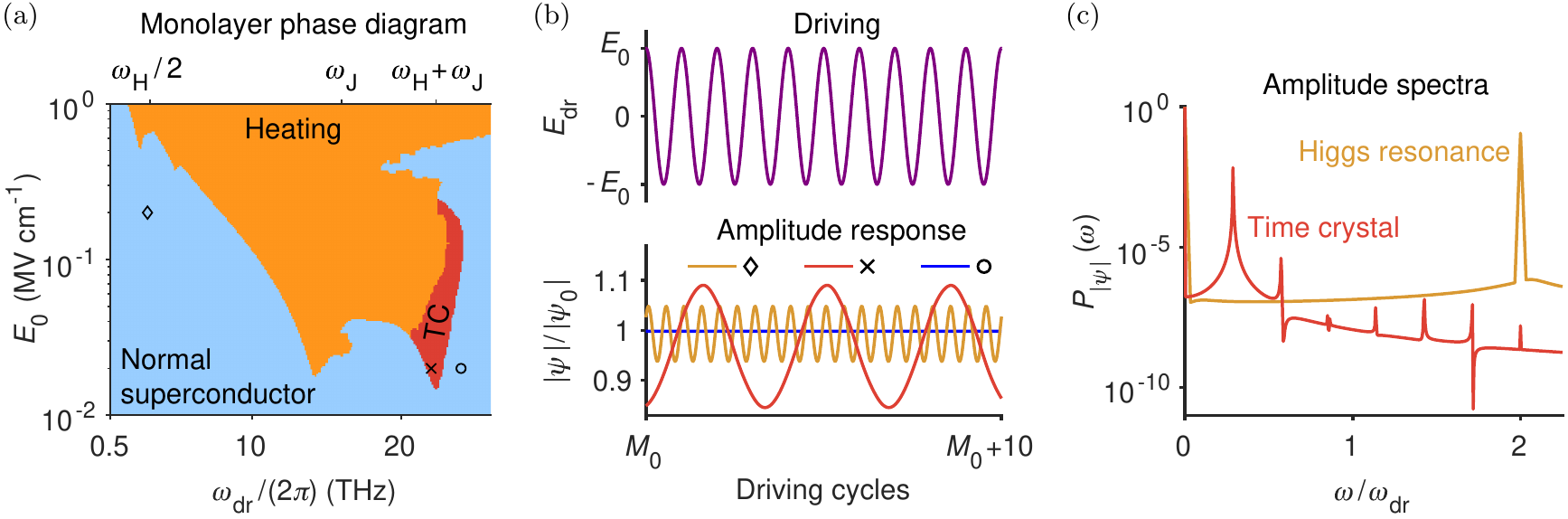}
	\caption{Dynamical phases of a light-driven monolayer cuprate superconductor. (a) Dynamical phase diagram of a monolayer cuprate continuously driven by an electric field with frequency $\omega_\mathrm{dr}$ and effective field strength $E_0$ at $T=0$. The time crystalline (TC) phase is encoded in red. (b) Driving $E_{\mathrm{dr}}(t)$ and response of the condensate amplitude $|\psi|/|\psi_0|(t)$ for the Higgs resonance at $\omega_{\mathrm{dr}}= \omega_{\mathrm{H}}/2$ (diamond), the time crystal (cross), and an off-resonantly driven superconductor (circle). The driving parameters are indicated by the symbols in (a). (c) Power spectra of the condensate amplitude, corresponding to the trajectories of the Higgs resonance and the time crystal presented in (b). The parameters for the monolayer system are the same as in Fig.~\ref{fig:1}.}
	\label{fig:2} 
\end{figure*}

To identify the Higgs time crystalline phase, we map out the dynamical phase diagram of optically driven high-$T_c$ superconductors as a function of the driving frequency $\omega_{\mathrm{dr}}$ and the driving amplitude $E_0$, which is shown in Fig.~\ref{fig:2}(a), for instance. The time crystalline state is induced by driving the superconductor via the non-linear coupling $\sim a^2 h$ of the electromagnetic field $a$ and the Higgs field $h$.  We demonstrate that driving at the frequency $\omega_{\mathrm{dr}} = \omega_{\mathrm{H}} + \omega_{\mathrm{J}}$ induces a time crystalline phase, where $\omega_{\mathrm{H}}$ is the Higgs frequency and $\omega_{\mathrm{J}}$ is the plasma frequency, as depicted in Fig.~\ref{fig:1}(d). We note that this non-linear coupling has been confirmed in conventional superconductors \cite{Matsunaga2014, Tsuji2015, Nakamura2019, Shimano2020}, while a direct probe of the Higgs field is challenging due to its scalar nature. Further studies on the Higgs mode in high-$T_c$ cuprates and organic superconductors are reported in Refs.~\cite{Peronaci2015, Katsumi2018, Buzzi2019, Chu2020, Schwarz2020, Puviani2020, Yang2020}. Persistent multi-frequency dynamics of the superconducting order parameter has been investigated in Ref.~\cite{Yuzbashyan2006}.

To describe the dynamics of optically driven superconductors, we develop a lattice gauge simulation that describes the motion of the order parameter of the superconducting state $\psi(\mathbf{r},t)$ and the electromagnetic field $\mathbf{A}(\mathbf{r},t)$. We first utilise our method to show how to induce the time crystalline state, and to determine its regime in the dynamical phase diagram. Furthermore, we demonstrate the robustness of the time crystalline phase against thermal fluctuations, and show that it can be realised and identified under pulsed operation.

\section{Three-dimensional lattice gauge model}
We represent the layered structure of high-$T_c$ superconductors via the lattice geometry illustrated in Fig.~\ref{fig:1}(b). We note that this geometry of CuO$_2$ layers perpendicular to the $c$-axis has motivated a low-energy description of stacks of Josephson junctions \cite{Koyama1996, Marel2001, Koyama2002}, which captures the appearance of Josephson plasma excitations reported in Refs.~\cite{Shibata1998, Dulic2001, Basov2005}.
Each layer is represented by a square lattice, leading to a discretisation of the fields of the form \mbox{$\psi(\mathbf{r},t) \to \psi_{l,m,n}(t) \equiv \psi_{\mathbf{r}}(t)$}. The in-plane discretisation length $d_{ab}$ constitutes a short-range cut-off well below the in-plane coherence length.
In doing so, we generalise the modelling of layered cuprates to a three-dimensional (3D) lattice of Josephson junctions. Each component of the vector potential $A_{i,\mathbf{r}}(t)$ is located between a lattice site $\mathbf{r}$ and its nearest neighbour $\mathbf{r}'(i)$ in the $i$-direction, where $i \in \{x,y,z\}$. According to the Peierls substitution, it describes the averaged electric field along the bond of a plaquette in Fig.~\ref{fig:1}(b).

We focus on temperatures below $T_c$, where the dominant low-energy degrees of freedom are Cooper pairs. We describe the Cooper pairs as a condensate of interacting bosons with charge $-2e$, represented by the complex field $\psi_{\mathbf{r}}(t)$. To construct the Hamiltonian of the lattice gauge model, we discretise the Ginzburg-Landau free energy \cite{Ginzburg1950} on a layered lattice and add time-dependent terms. We explicitly simulate the coupled dynamics of the condensate and the electromagnetic field. We discretise space by mapping it on a lattice, as mentioned, but implement the compact $U(1)$ lattice gauge theory in the time continuum limit \cite{Kogut1979}.
The particle-hole symmetry inherent to our relativistic model creates stable Higgs oscillations, even in bilayer cuprates where the Higgs frequency is between the two longitudinal Josephson plasma frequencies.

We consider mono- and bilayer cuprate superconductors. For bilayer cuprates, we assign the strong (weak) junctions to the even (odd) layers. The corresponding tunnelling coefficients are $t_{2n}=t_s$ and $t_{2n+1}=t_w$. The interlayer spacings $d_{2n,2n+1}=d_{s,w}$ are the distances between the CuO$_2$ planes in the crystal. Note that we suppose the $z$-direction to be aligned with the $c$-axis of the crystal throughout this paper.
The Hamiltonian of the lattice gauge model is
\begin{equation} \label{eq:Hamiltonian}
\mathcal{H} = \mathcal{H}_{\mathrm{sc}} + \mathcal{H}_{\mathrm{em}} + \mathcal{H}_{\mathrm{kin}} .
\end{equation}
The first term is the $|\psi|^4$ model of the superconducting condensate in the absence of Cooper pair tunnelling:
\begin{equation}\label{eq:HSC}
\mathcal{H}_{\mathrm{sc}} = \sum_{\mathbf{r}} \frac{| \pi_{\mathbf{r}} |^2}{K \hbar^2} - \mu | \psi_{\mathbf{r}} |^2 + \frac{g}{2} | \psi_{\mathbf{r}} |^4 ,
\end{equation}
where $\pi_{\mathbf{r}}= K \hbar^2 \partial_{t} \psi_{\mathbf{r}}^*$ is the conjugate momentum of $\psi_{\mathbf{r}}$, $\mu$ is the chemical potential, and $g$ is the interaction strength. This Hamiltonian is particle-hole symmetric due to its invariance under $\psi_{\mathbf{r}} \rightarrow \psi_{\mathbf{r}}^*$. 
The coefficient $K$ describes the magnitude of the dynamical term.

The electromagnetic part $\mathcal{H}_{\mathrm{em}}$ is the discretised form of the free field Hamiltonian, modified by tunable interlayer permittivities $\epsilon_{s,w}$ to capture the screening due to bound charges in the material:
\begin{equation} \label{eq:HEM}
\mathcal{H}_{\mathrm{em}} = \sum_{i,\mathbf{r}} \frac{\kappa_{i,\mathbf{r}} \epsilon_{i,\mathbf{r}} \epsilon_0}{2} E_{i,\mathbf{r}}^2 + \frac{\kappa_{z,\mathbf{r}}}{\kappa_{i,\mathbf{r}} \beta_{i,\mathbf{r}}^2 \mu_0} \Bigl[1 - \cos\bigl(\beta_{i,\mathbf{r}} B_{i,\mathbf{r}} \bigr) \Bigr] ,
\end{equation}
where $E_{i,\mathbf{r}}$ denotes the $i$-component of the electric field. The vector potential is located on the bonds between the superconducting sites. Consequently, this applies to the electric field as well. Note that we choose the temporal gauge for our calculations, i.e., \mbox{$E_{i,\mathbf{r}} = -\partial_{t} A_{i,\mathbf{r}}$}. Meanwhile, the magnetic field components \mbox{$B_{i,\mathbf{r}}= \epsilon_{ijk}\delta_j A_{k,\mathbf{r}}$} are centred about the plaquettes. This arrangement is consistent with the finite-difference time-domain (FDTD) method for solving Maxwell's equations \cite{Yee1966}. We calculate the spatial derivatives according to $\delta_j A_{k,\mathbf{r}} = (A_{k,\mathbf{r}'(j)}-A_{k,\mathbf{r}})/d_{j,\mathbf{r}}$, where $d_{j,\mathbf{r}}$ is the length of the bond.
The dielectric permittivities are $\epsilon_{x,\mathbf{r}}= \epsilon_{y,\mathbf{r}}= 1$ and $\epsilon_{z,\mathbf{r}}= \epsilon_n$. The other prefactors in Eq.~\eqref{eq:HEM} account for the anisotropic lattice geometry. They are defined as \mbox{$\kappa_{x,\mathbf{r}}= \kappa_{y,\mathbf{r}}= 1$} and $\kappa_{z,\mathbf{r}}= d_n/d_c$, while $\beta_{x,\mathbf{r}}= \beta_{y,\mathbf{r}}= 2ed_{ab}d_{n}/\hbar$ and $\beta_{z,\mathbf{r}}= 2ed_{ab}^2/\hbar$, where $d_c = (d_s + d_w)/2$.

The non-linear coupling between the Higgs field and the electromagnetic field derives from the tunnelling term
\begin{equation}
\mathcal{H}_{\mathrm{kin}} = \sum_{i,\mathbf{r}} t_{i,\mathbf{r}} |\psi_{\mathbf{r'}(i)} - \psi_{\mathbf{r}} \mathrm{e}^{\mathrm{i} a_{i,\mathbf{r}}}|^2 .
\end{equation}
 The unitless vector potential \mbox{$a_{i,\mathbf{r}}= -2e d_{i,\mathbf{r}} A_{i,\mathbf{r}}/\hbar$} couples to the phase of the superconducting field, ensuring the gauge-invariance of $\mathcal{H}_{\mathrm{kin}}$. The in-plane tunnelling coefficient is $t_{ab}$, and the $c$-axis tunnelling coefficients are $t_{s,w}$.

We solve the equations of motion for $\psi_{\mathbf{r}}(t)$ and $\mathbf{A}_{\mathbf{r}}(t)$ obtained from the Hamiltonian numerically, employing Heun's method with an integration step size $\Delta t= 1.6~\mathrm{as}$.
Thermal fluctuations are included by adding dissipation and Langevin noise to the equations of motion for both fields. For example, the time evolution of the superconducting field is given by
\begin{equation}
\partial_{t} \pi_{\mathbf{r}} = - \frac{\partial \mathcal{H}}{\partial \psi_{\mathbf{r}}} - \gamma \pi_{\mathbf{r}} + \xi_{\mathbf{r}} ,
\end{equation}
where $\gamma$ is a damping constant and $\xi_{\mathbf{r}}$ represents white Gaussian noise with zero mean, see Table~\ref{tab:parameters} for noise correlations. 
We note that the inclusion of in-plane dynamics and arbitrarily strong amplitude fluctuations constitutes a qualitative advance of previous descriptions, such as 1D sine-Gordon models \cite{Denny2015, Okamoto2016}.

We determine the response of the superconductor to periodic driving of the electric field along the $c$-axis. The external drive $E_{\mathrm{dr}}(t)$ has the frequency $\omega_{\mathrm{dr}}$ and the effective field strength $E_0$. We consider the long-wavelength limit such that the external drive is assumed to be homogeneous in the bulk of the sample. Thus, the time evolution of $E_{z,\mathbf{r}}(t)$ reads
\begin{equation} \label{eq:drive}
\partial_{t} {E}_{z,\mathbf{r}} = \frac{d_c}{d_n \epsilon_n \epsilon_0} \frac{\partial \mathcal{H}}{\partial A_{z,\mathbf{r}}} - \gamma E_{z,\mathbf{r}} + \eta_{z,\mathbf{r}} + \frac{\partial_{t} {E}_{\mathrm{dr}}}{\epsilon_n} ,
\end{equation}
where $\eta_{z,\mathbf{r}}$ is white Gaussian noise with zero mean. The equations of motion for $E_{x,\mathbf{r}}(t)$ and $E_{y,\mathbf{r}}(t)$ are analogous to Eq.~\eqref{eq:drive}, except for the driving term. We characterise the response by evaluating the sample averages of the condensate amplitude $|\psi(t)|$ and the supercurrent density $J(t)$, see also Appendix~\ref{app:observables}.

By applying the optical driving as described, we obtain the full dynamical phase diagram due to the direct coupling of the electromagnetic field to the superconducting order parameter. We note that resonant optical driving of phonon modes has been utilised and discussed in Refs.~\cite{Fausti2011, Hu2014, Cremin2019, Denny2015, Okamoto2016, Okamoto2017}. Here, we ignore the phononic resonances, so that our predictions are valid away from these resonances. A combined description will be given elsewhere.

\section{Two-mode model}
Before we present the full numerical simulation, we identify the main resonant phenomena of the system. We consider the zero-temperature limit, where the in-plane dynamics can be neglected and the model simplifies to a 1D chain along the $c$-axis. Furthermore, we restrict ourselves to weak driving and a monolayer structure with $t_s=t_w \equiv t_{\mathrm{J}}$ and $d_s=d_w \equiv d$. For periodic boundary conditions, the time evolution then reduces to two coupled equations of motion. 
Keeping only linear terms except for the lowest order coupling between the Higgs field and the unitless vector potential, we find
\begin{align}
	\partial_{t}^2 a + \gamma \partial_{t} a + \omega_{\mathrm{J}}^2 a + 2\omega_{\mathrm{J}}^2 ah &\approx j_{\mathrm{dr}} , \label{eq:eoma} \\
	\partial_{t}^2 h + \gamma \partial_{t} h + \omega_{\mathrm{H}}^2 h + \alpha \omega_{\mathrm{J}}^2 a^2 &\approx 0 , \label{eq:eomh}
\end{align}
where $h = (\psi-\psi_0)/\psi_0$ is the Higgs field with $\psi_0$ being the equilibrium condensate amplitude, $\gamma$ is the damping constant, and $\alpha$ is the capacitive coupling constant of the junction. Note that the unitless vector potential $a$ equals the phase difference between adjacent planes in this setting. The external drive appears through the current $j_{\mathrm{dr}}$.
The Higgs and plasma frequencies are $\omega_{\mathrm{H}}=\sqrt{2\mu/K \hbar^2}$ and $\omega_{\mathrm{J}}=\sqrt{t_{\mathrm{J}}/\alpha K \hbar^2}$, respectively.

The main finding of this work is the emergence of a time crystalline phase by driving at the sum of the Higgs and plasma frequencies, $\omega_{\mathrm{dr}}=\omega_{\mathrm{J}} + \omega_{\mathrm{H}}$. A cubic interaction process, visualised in Fig.~\ref{fig:1}(d), allows for simultaneous resonant driving of both the Higgs and the plasma modes \cite{Supp}.

In addition to the sum resonance, we identify various other resonances from the simplified equations of motion. For a response of the vector potential at the driving frequency, i.e., $a=a_1 \mathrm{cos}( \omega_{\mathrm{dr}}t)$, Eq.~\eqref{eq:eomh} simplifies to a forced oscillator with a resonance at $\omega_{\mathrm{dr}}=\omega_{\mathrm{H}}/2$. This recovers the sub-gap Higgs resonance \cite{Shimano2020}. The sub-gap resonance and the sum resonance originate from the same cubic coupling term $\sim a^2 h$, as illustrated in Figs.~\ref{fig:1}(c) and \ref{fig:1}(d).
Next, we consider the range of driving frequencies where the Higgs field exhibits a second-harmonic response, that is, the external drive induces Higgs oscillations of the form $h = h_0 + h_1\mathrm{cos}( 2\omega_{\mathrm{dr}}t)$ through the $a^2$ term in Eq.~\eqref{eq:eomh}. For small driving amplitudes, the $ah$ term in Eq.~\eqref{eq:eoma} can be neglected so that the equation reduces to a forced oscillator with a resonance at $\omega_{\mathrm{dr}}= \omega_{\mathrm{J}}$. However, the response is modified once the coupling to the Higgs field becomes significant. Then, Eq.~\eqref{eq:eoma} approaches a parametrically driven oscillator. The parametric resonances emerge at $\omega_{\mathrm{dr}}= \omega_{\mathrm{J}}/k$, where $k \in \mathbb{N}$.

\section{Dynamical phase diagram}
We now present our numerical results in two steps. Firstly, we verify our analytical predictions for the resonances and, in particular, the Higgs time crystal by mapping out the dynamical phase diagrams of mono- and bilayer cuprate superconductors at zero temperature. We will show how the sum resonance is modified in a bilayer system, which has two plasma modes. Secondly, we test the robustness of this phase against thermal fluctuations using finite-temperature simulations.

\subsection{Monolayer cuprate superconductor}
Here, we consider a monolayer cuprate with \mbox{$\omega_{\mathrm{H}}/2 \pi \approx6.3~\mathrm{THz}$}, $\omega_{\mathrm{J}}/2 \pi \approx16.0~\mathrm{THz}$, $\gamma/2 \pi= 0.5~\mathrm{THz}$, and $\alpha= 0.33$, see Table~\ref{tab:parameters} for full parameter set.
The system is continuously driven at various amplitudes and frequencies in the terahertz regime. In each realisation, the drive is applied for 20 ps and the relevant frequency spectra are computed using the final 10 ps, which amounts to $5 < M_{\mathrm{tot}} < 300$ driving cycles in the frequency range of interest. The dynamical phase diagram in Fig.~\ref{fig:2}(a) is mapped out by analysing the normalised power spectra of $|\psi(t)|$ and $J(t)$ defined as $P_f(\omega) = \langle f(\omega)f(-\omega) \rangle$, where $\int P_f(\omega)  d\omega = 1$, $f(\omega)=1/\sqrt{T_s} \int dt' \mathrm{exp}(-\mathrm{i}\omega t') f(t')$, and $T_s=10$ ps is the sampling interval. 
Specifically, we obtain the spectral entropy for the dynamics of the condensate amplitude, $\mathcal{S}_{|\psi|} = - \int  d\omega P_{|\psi|}(\omega) \mathrm{ln} P_{|\psi|}(\omega)$.

The heating regime, which is characterised by a strong depletion of the condensate, is identified based on the threshold $\mathcal{S}_{|\psi|}>2.2\times 10^{-2}$. It indicates the appearance of resonant phases associated with the Higgs and plasma excitations.
We note that the two dominant heating tongues are weakly red-detuned from the expected resonance frequencies $\omega_{\mathrm{H}}/2$ and $\omega_{\mathrm{J}}$, respectively. Such a renormalisation of the fundamental frequencies is inherent to strongly driven non-linear systems \cite{Landau1976}. This effect is further amplified by the damping terms present in our model. We identify the small tongue at $\omega_{\mathrm{dr}}/2 \pi \approx 4.8$ THz as the third order parametric resonance of the Josephson plasma mode around $\omega_{\mathrm{J}}/3$.

For intermediate driving intensity, we observe several dynamical regimes due to resonances.
The resonance with the lowest frequency is the Higgs resonance at $\omega_{\mathrm{dr}} = \omega_{\mathrm{H}}/2$.
In general, resonant excitation of the Higgs mode is marked by strong modulation of the condensate amplitude as exemplified in Fig.~\ref{fig:2}(b). Moreover, the Higgs resonance exhibits a commensurate and superharmonic response of $|\psi(t)|$ with respect to the driving $E_{\mathrm{dr}}(t)$ as seen from the closed trajectory in Fig.~\ref{fig:1}(c) and the sharp peak at $2\omega_{\mathrm{dr}}$ of the condensate amplitude spectrum in Fig.~\ref{fig:2}(c). We emphasise that driving away from any noticeable resonance, indicated as the blue regime in Fig.~\ref{fig:2}(a), induces only a single sharp peak in the supercurrent spectrum, namely at the driving frequency. The condensate amplitude oscillates at twice the driving frequency in the blue regime. This also applies to the regime near the Josephson plasma resonance at $\omega_{\mathrm{dr}} = \omega_{\mathrm{J}}$, where the system responds with strong oscillations of the supercurrent.

The red regime in Fig.~\ref{fig:2}(a), identified via the condition \mbox{$10^{-4}<\mathcal{S}_{|\psi|}<2.2\times 10^{-2}$}, is the Higgs time crystal introduced earlier. We emphasise that its resonance condition $\omega_{\mathrm{dr}}=\omega_{\mathrm{J}}+\omega_{\mathrm{H}}$ differs from the sub-gap frequencies $\omega_{\mathrm{dr}} \lesssim \omega_{\mathrm{H}}/2$ used in standard Higgs spectroscopy. 
The sum resonance simultaneously couples to the Higgs and plasma resonances as evident from the exemplary mean-field trajectory in Fig.~\ref{fig:1}(d), where the amplitude oscillation is accompanied by a strong oscillation of the phase difference between the junctions. Despite a smaller driving amplitude $E_0$, the plasma mode is excited with larger amplitude than for the Higgs resonance. The strong activation of the plasma mode results in a partial depletion of the condensate as visible in Fig.~\ref{fig:2}(b), where the time average of the oscillatory motion of the condensate amplitude is below 1.
The key feature of the novel phase is the subharmonic response of the condensate amplitude as $|\psi(t)|$ oscillates at $\omega_\mathrm{H}$ when the superconductor is driven at $\omega_{\mathrm{dr}}=\omega_{\mathrm{J}}+\omega_{\mathrm{H}}$. This phenomenon is highlighted in Fig.~\ref{fig:2}(b) and in the strong subharmonic peak in the power spectrum of $|\psi(t)|$ shown in Fig.~\ref{fig:2}(c). 
The other dynamical phases respect the time-translation symmetry imposed by the external drive as evidenced by Figs.~\ref{fig:2}(b) and \ref{fig:2}(c).

The subharmonic collective motion is one of the defining features of a time crystal. In addition to being subharmonic, the response of the time crystalline state is also incommensurate to the external driving. That is, the phase-space trajectory traces an open loop for any number of driving cycles, see also Fig.~\ref{fig:1}(d).
Therefore, and more specifically, the state that we propose to create is an incommensurate time crystal in high-$T_c$ superconductors. We will confirm its robustness against perturbations of the drive and thermal fluctuations for the bilayer case. We note that the subharmonic response can be expected to be rigid as it emerges for a broad regime of driving parameters rather than a fine-tuned point in the dynamical phase diagram. In addition, our finite-temperature calculations with a large number of lattice sites will highlight the many-body nature of the Higgs time crystal.

\begin{figure}[!t]
	\centering
	\includegraphics[scale=1]{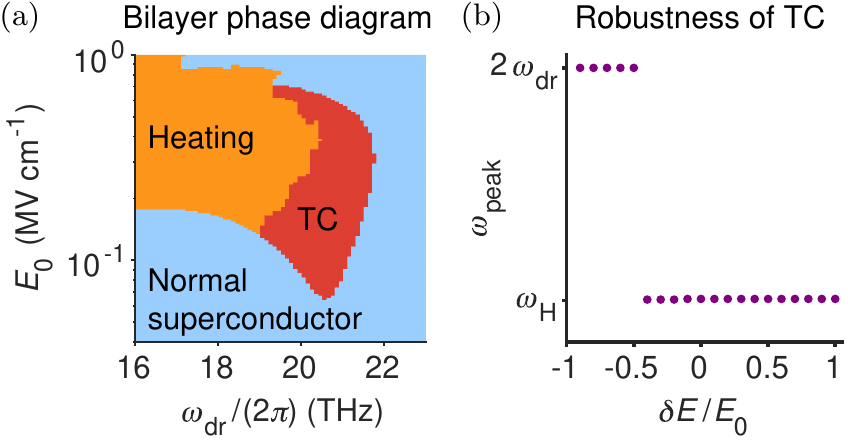}
	\caption{Higgs time crystal in a light-driven bilayer cuprate superconductor. (a) Dynamical phase diagram of a bilayer cuprate continuously driven by an electric field with frequency $\omega_\mathrm{dr}$ and effective field strength $E_0$ at $T=0$. (b) Robustness of the time crystal (TC) against perturbations of the drive as described in the text. Values of the dominant amplitude frequency $\omega_{\mathrm{peak}}$ close to $\omega_{\mathrm{H}}$ indicate a subharmonic response, whereas maxima at $2\omega_{\mathrm{dr}}$ mark a normal response. The bilayer system has the Higgs frequency $\omega_{\mathrm{H}}/2\pi \approx6.3$ THz and the two longitudinal Josephson plasma frequencies $\omega_{\mathrm{J}1}/2\pi \approx2.0~\mathrm{THz}$ and $\omega_{\mathrm{J}2}/2\pi \approx14.3~\mathrm{THz}$ at $T=0$, see Table~\ref{tab:parameters} for full parameter set.}
	\label{fig:3} 
\end{figure}

\subsection{Bilayer superconductor}
We now focus on bilayer cuprates. Due to the staggered tunneling coefficients $t_s$ and $t_w$ along the $c$-axis, the system has two fundamental longitudinal plasma excitations with frequencies $\omega_{\mathrm{J1}}$ and $\omega_{\mathrm{J2}}$.
The dynamical phase diagram at zero temperature in Fig.~\ref{fig:3}(a) displays a regime, in which a Higgs time crystal is induced by optical driving at a sum resonance. Here, the resonance condition is $\omega_{\mathrm{dr}}=\omega_{\mathrm{H}}+\omega_{\mathrm{J2}}$, so it is the sum of the Higgs frequency and the upper plasma frequency.

First, we examine how perturbing the optical drive itself affects the subharmonic response. To excite the sum resonance, we initially drive the bilayer superconductor with $E_0= 0.1~\mathrm{MV \, cm^{-1}}$ and $\omega_{\mathrm{dr}}/2\pi= 21~\mathrm{THz}$. At some instant of time $t_0$, the driving is altered so that the oscillation amplitude of the field strength depends on its sign for $t>t_0$:
\begin{equation}
E_{\mathrm{dr}}(t)= 
\begin{cases}
E_0 \mathrm{cos}(\omega_{\mathrm{dr}} t) & \mathrm{for}~\mathrm{cos}(\omega_{\mathrm{dr}} t) \geq 0 , \\
(E_0 + \delta E) \, \mathrm{cos}(\omega_{\mathrm{dr}} t) & \mathrm{for}~\mathrm{cos}(\omega_{\mathrm{dr}} t) < 0 .
\end{cases}
\end{equation}
After allowing the system to relax to a steady state, we take the power spectrum of the condensate amplitude and determine the dominant frequency $\omega_{\mathrm{peak}}$.
The robustness of the subharmonic response is demonstrated by Fig.~\ref{fig:3}(b), where perturbations of the driving amplitude between $\delta E/E_0= -0.4$ and $\delta E/E_0= 1$ do not destroy the sum resonance.
We have also verified the persistence of the subharmonic response for $10^5$ cycles of continuous driving at $T=0$ \cite{Supp}. Because of experimental and numerical limitations in accessible timescales ($\sim 10^2$ driving cycles for our finite-temperature calculations), we will not distinguish here between a `true' time crystal and a slowly decaying time crystal \cite{Choi2017, Yao2020}.

\begin{figure*}[!t]
	\centering
	\includegraphics[scale=1]{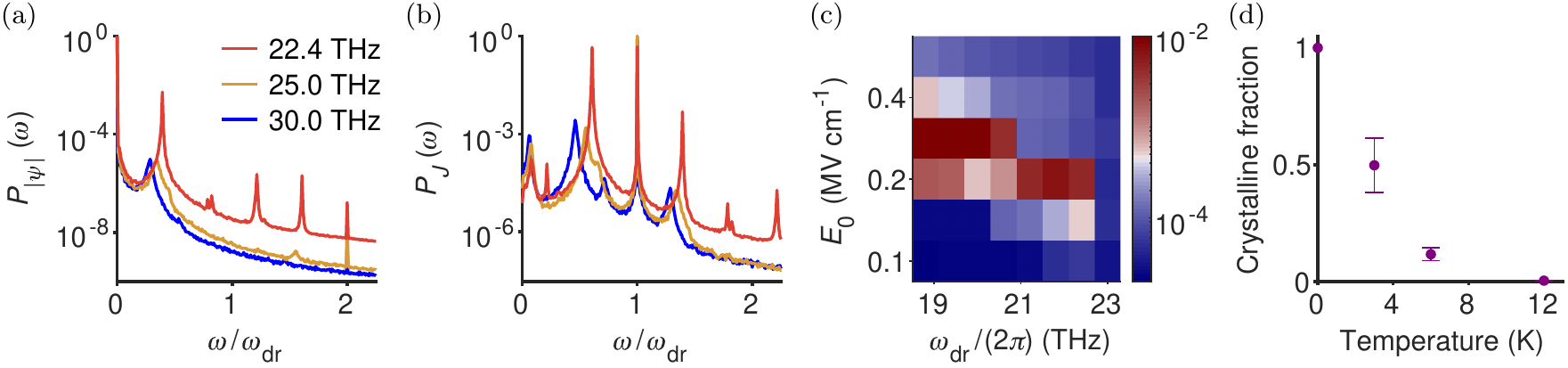}
	\caption{Higgs time crystal at non-zero temperatures. (a) Comparison between the power spectra of the condensate amplitude at $T=3~\mathrm{K}~\sim 0.1 \, T_c$ for $E_0= 0.2~\mathrm{MV \, cm^{-1}}$ and different driving frequencies indicated in the legend. The time crystalline state at $\omega_{\mathrm{dr}}/2\pi = 22.4~\mathrm{THz}$ is demonstrated by the strongly enhanced subharmonic peak at $\omega_{\mathrm{H}}$. (b) Power spectra of the supercurrent density for the same parameters as in (a). The time crystalline state creates strongly enhanced side peaks at $\omega_{\mathrm{dr}} \pm \omega_{\mathrm{H}}$. (c) Time crystalline fraction $P_{J}(\omega_{\mathrm{dr}} + \omega_{\mathrm{H}})$ in a section of the dynamical phase diagram at \mbox{$T=3~\mathrm{K}~\sim 0.1 \, T_c$}, containing the time crystalline phase. (d) Temperature dependence of the optimal time crystalline fraction for a bilayer cuprate superconductor, rescaled by its value at $T=0$. The optimal crystalline fraction at a given temperature corresponds to the maximum value of $P_{J}(\omega_{\mathrm{dr}} + \omega_{\mathrm{H}})$ in the relevant section of the dynamical phase diagram, as exemplified in (c). The error bars in (d) arise from the standard errors of Lorentzian fits to the blue-detuned side peaks. The parameters for the bilayer system are the same as in Fig.~\ref{fig:3}. The resonance frequencies are shifted at finite temperature.}
	\label{fig:4} 
\end{figure*}

We note that the time crystalline response is stabilised by the non-linear coupling between the Higgs and plasma modes, which further highlights the collective nature of the Higgs time crystal.
Furthermore, the amplitudes of the oscillations are saturated by non-linear processes in the system, see Ref. \cite{Nayfeh1982} for example, while the dissipative coupling to the environment limits heating.

Next, we demonstrate the robustness of the Higgs time crystal against thermal fluctuations modelled as Langevin noise in the dynamics of the fields. These fluctuations are a natural test for the rigidity of the subharmonic response against temporal perturbations \cite{Yao2020}.
When considering thermal fluctuations, we include the in-plane dynamics of the fields in a full 3D simulation. The complete parameter set is summarised in Table~\ref{tab:parameters}, implying the Higgs frequency \mbox{$\omega_{\mathrm{H}}/2\pi \approx6.3~\mathrm{THz}$} and the two longitudinal Josephson plasma frequencies \mbox{$\omega_{\mathrm{J}1}/2\pi \approx2.0~\mathrm{THz}$} and \mbox{$\omega_{\mathrm{J}2}/2\pi \approx14.3~\mathrm{THz}$} at $T=0$. For simplicity, we keep the chemical potential fixed in the following finite-temperature calculations, $\mu(T) \equiv \mu$.
We choose the parameters within the CuO$_2$ planes to yield a critical temperature of $T_c \sim 30~\mathrm{K}$. We find that a discretisation of $48\times 48 \times 4$ lattice sites with periodic boundaries is sufficient to obtain fully converged results with respect to the system size. Note that both the Higgs and Josephson plasma frequencies are renormalised at finite temperature \cite{Supp}.

Examples of the power spectra of the condensate amplitude and the supercurrent density at $T=3~\mathrm{K}$ are shown in Figs.~\ref{fig:4}(a) and \ref{fig:4}(b), respectively. 
When the sum resonance is driven, the condensate amplitude exhibits strong subharmonic modulation as evidenced by a sharp peak in the amplitude spectrum in Fig.~\ref{fig:4}(a). Moreover, we observe in Fig.~\ref{fig:4}(a) how the modulation of the condensate amplitude is suppressed as the driving frequency is tuned away from the resonance frequency.
As shown in Fig.~\ref{fig:4}(b), we identify an experimentally relevant signature of the superconducting time crystalline phase, which is the appearance of two side peaks at $\omega_{\mathrm{dr}} \pm \omega_{\mathrm{H}}$ in the power spectrum of the supercurrent density. The side peaks vanish as the driving frequency is tuned away from the resonance frequency. Coherent dynamics of supercurrents can be experimentally probed using second-harmonic measurements \cite{vonHoegen2018,vonHoegen2019}.

To quantify the time crystalline fraction, we use the height of the blue-detuned side peak in the power spectrum of the supercurrent density, $P_{J}(\omega_{\mathrm{dr}} + \omega_{\mathrm{H}})$.  Figure~\ref{fig:4}(d) displays the temperature dependence of the optimal crystalline fraction for a bilayer cuprate, normalised to the optimal time crystalline fraction at $T=0$. The optimal driving parameters at each temperature were inferred from coarse scans such as that in Fig.~\ref{fig:4}(c).
As we expect for time crystals under increasingly strong perturbation, the crystalline fraction decreases with temperature. Nevertheless, the subharmonic response is robust against thermal noise for temperatures up to \mbox{$T=6~\mathrm{K} \sim 0.2 \, T_c$}.

\section{Pulsed excitation of the\\Higgs time crystal}
\begin{figure}[!t]
	\centering
	\includegraphics[scale=1]{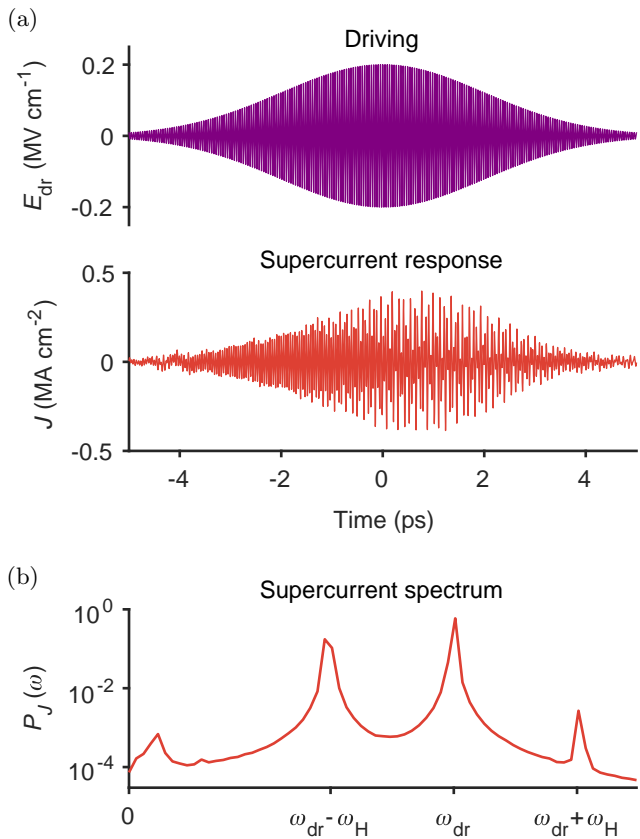}
	\caption{Time crystalline response of a bilayer cuprate superconductor to a driving pulse. (a) Temporal waveform of the pulsed electric field and the induced motion of the supercurrent density shown for one representative trajectory at $T=3~\mathrm{K}~\sim 0.1 \, T_c$ with an effective field strength $E_0=0.2~\mathrm{MV~cm}^{-1}$, driving frequency $\omega_{\mathrm{dr}}/2\pi=22.4~\mathrm{THz}$, and pulse width $\sigma=2~\mathrm{ps}$. (b) Power spectrum of the supercurrent density, measured in the interval between 0 and 2~ps.  The parameters for the bilayer system are the same as in Fig.~\ref{fig:3}.}
	\label{fig:5} 
\end{figure}

While significant progress has been made in generating continuous-wave terahertz sources \cite{Welp2013}, typical experiments in optically driven superconductors utilise pulsed excitation, as in most pump-probe experiments. We now point out that the time crystalline phase can be detected when the system is driven with a short pulse, rather than the steady state discussed so far.
We consider a pulsed driving scheme by introducing a Gaussian envelope of the periodic driving, that is, \mbox{$E_{\mathrm{dr}}(t)= E_0 \mathrm{cos}(\omega_{\mathrm{dr}} t) \, \mathrm{exp}(-t^2/2 \sigma^2)$} with the pulse width $\sigma$. In Fig.~\ref{fig:5}, we present an example of the dynamical response of the bilayer system under pulsed excitation.
The response shown in Fig.~\ref{fig:5}(b) is approximately the Fourier broadened form of Fig.~\ref{fig:4}(b). The similarity between the two results suggests that the defining features of the Higgs time crystal of continuously driven superconductors are detectable for pulsed driving protocols with realistic pulse lengths. The response can be clearly distinguished from normal dynamical phases by probing the coherent dynamics of the supercurrent.
Thus, the Higgs time crystal predicted here can be observed in current state-of-the-art experiments with optically driven high-$T_c$ superconductors.

\section{Discussion}
In conclusion, we have demonstrated the emergence of a time-crystalline phase in a high-$T_c$ superconductor, which is induced by optical driving of a sum resonance of the Higgs mode and a Josephson plasma mode. Using a newly developed lattice gauge simulator, we demonstrate this time crystal for mono- and bilayer cuprates, and show its robustness against thermal fluctuations, for up to $\sim 20\%$ of the critical temperature. As an experimentally accessible signature we observe the emergence of two side peaks at $\omega_{\mathrm{dr}} \pm \omega_{\mathrm{H}}$ in the supercurrent spectra. This signature is also visible in pulsed operation, which mimics realistic experimental conditions.

The emergent time crystalline order that we propose to induce, constitutes a qualitative departure from previous light-induced states in solids, because it is a genuine non-equilibrium state with no equilibrium counterpart. The realisation of such a state expands the scope of the scientific effort to design many-body states by optical driving beyond the paradigm of renormalising equilibrium orders. While even this existing paradigm has been and continues to be thought-provoking and stimulating, the work presented here urges the design and exploration of light-induced non-equilibrium states beyond that framework, and thereby expands the scope of the effort to design quantum matter on demand.

\begin{acknowledgments}
We thank Andrea Cavalleri, Junichi Okamoto, and Kazuma Nagao for fruitful discussions. This work is supported by the Deutsche Forschungsgemeinschaft (DFG) in the framework of SFB 925 and the Cluster of Excellence ``Advanced Imaging of Matter" (EXC 2056), Project No. 390715994.
\end{acknowledgments}

\appendix

\section{Noise correlations} \label{app:noise}
The fluctuation-dissipation theorem requires
\begin{align}
\langle \mathrm{Re} \{{\xi_{\mathbf{r}} (t)}\} \mathrm{Re} \{{\xi_{\mathbf{r'}} (t')}\} \rangle &= \frac{\gamma K \hbar^2 k_{\mathrm{B}} T}{V_0} \delta_{\mathbf{r}\mathbf{r'}} \delta(t-t') \, , \\
\langle \mathrm{Im} \{{\xi_{\mathbf{r}} (t)}\} \mathrm{Im} \{{\xi_{\mathbf{r'}} (t')}\} \rangle &= \frac{\gamma K \hbar^2 k_{\mathrm{B}} T}{V_0} \delta_{\mathbf{r}\mathbf{r'}} \delta(t-t') \, , \\
\langle \mathrm{Re} \{{\xi_{\mathbf{r}} (t)}\} \mathrm{Im} \{{\xi_{\mathbf{r'}} (t')}\} \rangle &= 0
\end{align}
for the noise term of the superconducting field, where $V_0= d_{ab}^2 d_c$ is the discretisation volume of a single superconducting site. The noise correlations for the electric field are
\begin{align}
\langle \eta_{x,\mathbf{r}} (t) \eta_{x,\mathbf{r'}} (t') \rangle &= \frac{2 \gamma k_{\mathrm{B}} T}{\epsilon_0 V_0} \delta_{\mathbf{r}\mathbf{r'}} \delta(t-t') , \\
\langle \eta_{y,\mathbf{r}} (t) \eta_{y,\mathbf{r'}} (t') \rangle &= \frac{2 \gamma k_{\mathrm{B}} T}{\epsilon_0 V_0} \delta_{\mathbf{r}\mathbf{r'}} \delta(t-t') , \\
\langle \eta_{z,\mathbf{r}} (t) \eta_{z,\mathbf{r'}} (t') \rangle &= \frac{d_c}{d_n \epsilon_n} \frac{2 \gamma k_{\mathrm{B}} T}{\epsilon_0 V_0} \delta_{\mathbf{r}\mathbf{r'}} \delta(t-t') .
\end{align}

\section{Characterisation of the response} \label{app:observables}
We characterise the response of the system to the periodic driving by studying the dynamics of the sample averages of the condensate amplitude and the supercurrent along the $c$-axis.
The supercurrent along a single junction in the $c$-direction is given by the Josephson relation 
\begin{equation}
J^z_{l,m,n} = \frac{4e t_n d_c}{\hbar}\mathrm{Im}\bigl\{\psi^*_{l,m,n+1}\psi_{l,m,n}e^{i a_{l,m,n}^z}\bigr\} .
\end{equation}
The sample average of the supercurrent density along the $c$-axis can be obtained from
\begin{equation}
J(t) = \frac{d_s \overline{J_s(t)} + d_w \overline{J_w(t)}}{d_s+d_w},
\end{equation}
where $\overline{J_{s,w}(t)}$ denotes the spatial average of Josephson currents along either strong or weak junctions. In the case of non-zero temperatures, we average the power spectra $P_{|\psi|}(\omega)$ and $P_{J}(\omega)$ over an ensemble of trajectories. We find that 100 trajectories are enough to obtain convergent results for sampling thermal fluctuations at non-zero temperatures.

\begin{table}[!h]
	\caption{Model parameters used in the simulations.}
	\renewcommand{\arraystretch}{1.5}
	\begin{tabular}{lrr}
		\hline
		& Monolayer & Bilayer \\
		\hline
		$K~(\text{meV}^{-1})$ & $\qquad 2.9 \times 10^{-5}$ & $\qquad 2.9 \times 10^{-5}$ \\
		$\mu~(\text{meV})$ & $1.0 \times 10^{-2}$ & $1.0 \times 10^{-2}$ \\
		$g~(\text{meV \AA}^3)$ & 5.0 & 5.0 \\
		$\gamma/2\pi~(\text{THz})$ & 0.5 & 0.5 \\
		\hline
		$t_{ab}~(\text{meV})$ & $6.2 \times 10^{-1}$ & $6.2 \times 10^{-1}$ \\
		$t_s~(\text{meV})$ & \multirow{2}{*}{$4.2 \times 10^{-2}$} &$2.5 \times 10^{-2}$ \\
		$t_w~(\text{meV})$ & &$1.0 \times 10^{-3}$ \\
		\hline
		$d_{ab}~(\text{\AA})$ & 15 & 15 \\
		$d_s~(\text{\AA})$ & \multirow{2}{*}{6} & 4  \\
		$d_w~(\text{\AA})$ &  & 8 \\
		\hline
		$\epsilon_s$ & \multirow{2}{*}{1} & 1 \\
		$\epsilon_w$ &  & 4 \\
		\hline
	\end{tabular}
	\renewcommand{\arraystretch}{1}
	\label{tab:parameters}
\end{table}

\section{Model parameters} \label{app:parameters}
Table~\ref{tab:parameters} summarises the parameters of our numerical calculations for mono- and bilayer systems, respectively. In both cases, our parameter choice of $\mu$ and $g$ implies an equilibrium condensate density \mbox{$n_0= \mu/g = 2 \times 10^{21}~\mathrm{cm^{-3}}$} at $T=0$. The bilayer system has two longitudinal $c$-axis plasma modes. Their eigenfrequencies are
\begin{widetext}
\begin{equation}
\omega_{\mathrm{J1,J2}}^2 = \biggl( \frac{1}{2}+ \alpha_{s} \biggr) \Omega_{s}^2 + \biggl( \frac{1}{2}+ \alpha_{w} \biggr) \Omega_{w}^2 \mp \sqrt{ \biggl[\biggl( \frac{1}{2}+ \alpha_{s} \biggr) \Omega_{s}^2 - \biggl( \frac{1}{2}+ \alpha_{w} \biggr) \Omega_{w}^2 \biggr]^2 + 4 \alpha_{s} \alpha_{w} \Omega_{s}^2 \Omega_{w}^2 } ,
\end{equation}
\end{widetext}
as follows from a sine-Gordon analysis at $T=0$ \cite{Marel2001, Koyama2002}. Here we introduced the bare plasma frequencies of the strong and weak junctions
\begin{equation}
\Omega_{s,w}= \sqrt{\frac{8 t_{s,w} n_0  e^2 d_c d_{s,w}}{\hbar^2 \epsilon_{s,w} \epsilon_0}} ,
\end{equation}
where $d_c= (d_s + d_w)/2$. The capacitive coupling constants are given by
\begin{equation}
\alpha_{s,w}= \frac{\epsilon_{s,w} \epsilon_0}{8 K n_0 e^2 d_c d_{s,w}} .
\end{equation}

\pagebreak
\noindent
Besides, there is a transverse $c$-axis plasma mode with the eigenfrequency
\begin{equation}
\omega_{\mathrm{T}}^2 = \frac{1+ 2\alpha_s + 2\alpha_w}{\alpha_s + \alpha_w} \, \Bigl( \alpha_{s} \Omega_s^2 + \alpha_w \Omega_w^2 \Bigr) .
\end{equation}
We have $\alpha_s=0.5$, $\alpha_w=1$, $\omega_{\mathrm{J1}}/2\pi \approx 2.0~\mathrm{THz}$, $\omega_{\mathrm{J2}}/2\pi \approx 14.3~\mathrm{THz}$, and $\omega_{\mathrm{T}}/2\pi \approx 11.8~\mathrm{THz}$ for the parameters specified in Table~\ref{tab:parameters}. The in-plane plasma frequency amounts to 154 THz.

\bibliography{biblio}

\providecommand{\noopsort}[1]{}\providecommand{\singleletter}[1]{#1}%
\begin{thebibliography}{48}%
\makeatletter
\providecommand \@ifxundefined [1]{%
 \@ifx{#1\undefined}
}%
\providecommand \@ifnum [1]{%
 \ifnum #1\expandafter \@firstoftwo
 \else \expandafter \@secondoftwo
 \fi
}%
\providecommand \@ifx [1]{%
 \ifx #1\expandafter \@firstoftwo
 \else \expandafter \@secondoftwo
 \fi
}%
\providecommand \natexlab [1]{#1}%
\providecommand \enquote  [1]{``#1''}%
\providecommand \bibnamefont  [1]{#1}%
\providecommand \bibfnamefont [1]{#1}%
\providecommand \citenamefont [1]{#1}%
\providecommand \href@noop [0]{\@secondoftwo}%
\providecommand \href [0]{\begingroup \@sanitize@url \@href}%
\providecommand \@href[1]{\@@startlink{#1}\@@href}%
\providecommand \@@href[1]{\endgroup#1\@@endlink}%
\providecommand \@sanitize@url [0]{\catcode `\\12\catcode `\$12\catcode
  `\&12\catcode `\#12\catcode `\^12\catcode `\_12\catcode `\%12\relax}%
\providecommand \@@startlink[1]{}%
\providecommand \@@endlink[0]{}%
\providecommand \url  [0]{\begingroup\@sanitize@url \@url }%
\providecommand \@url [1]{\endgroup\@href {#1}{\urlprefix }}%
\providecommand \urlprefix  [0]{URL }%
\providecommand \Eprint [0]{\href }%
\providecommand \doibase [0]{https://doi.org/}%
\providecommand \selectlanguage [0]{\@gobble}%
\providecommand \bibinfo  [0]{\@secondoftwo}%
\providecommand \bibfield  [0]{\@secondoftwo}%
\providecommand \translation [1]{[#1]}%
\providecommand \BibitemOpen [0]{}%
\providecommand \bibitemStop [0]{}%
\providecommand \bibitemNoStop [0]{.\EOS\space}%
\providecommand \EOS [0]{\spacefactor3000\relax}%
\providecommand \BibitemShut  [1]{\csname bibitem#1\endcsname}%
\let\auto@bib@innerbib\@empty
\bibitem [{\citenamefont {{Fausti}}\ \emph {et~al.}(2011)\citenamefont
  {{Fausti}}, \citenamefont {{Tobey}}, \citenamefont {{Dean}}, \citenamefont
  {{Kaiser}}, \citenamefont {{Dienst}}, \citenamefont {{Hoffmann}},
  \citenamefont {{Pyon}}, \citenamefont {{Takayama}}, \citenamefont
  {{Takagi}},\ and\ \citenamefont {{Cavalleri}}}]{Fausti2011}%
  \BibitemOpen
  \bibfield  {author} {\bibinfo {author} {\bibfnamefont {D.}~\bibnamefont
  {{Fausti}}}, \bibinfo {author} {\bibfnamefont {R.~I.}\ \bibnamefont
  {{Tobey}}}, \bibinfo {author} {\bibfnamefont {N.}~\bibnamefont {{Dean}}},
  \bibinfo {author} {\bibfnamefont {S.}~\bibnamefont {{Kaiser}}}, \bibinfo
  {author} {\bibfnamefont {A.}~\bibnamefont {{Dienst}}}, \bibinfo {author}
  {\bibfnamefont {M.~C.}\ \bibnamefont {{Hoffmann}}}, \bibinfo {author}
  {\bibfnamefont {S.}~\bibnamefont {{Pyon}}}, \bibinfo {author} {\bibfnamefont
  {T.}~\bibnamefont {{Takayama}}}, \bibinfo {author} {\bibfnamefont
  {H.}~\bibnamefont {{Takagi}}},\ and\ \bibinfo {author} {\bibfnamefont
  {A.}~\bibnamefont {{Cavalleri}}},\ }\bibfield  {title} {\bibinfo {title}
  {Light-induced superconductivity in a stripe-ordered cuprate},\ }\href
  {https://doi.org/10.1126/science.1197294} {\bibfield  {journal} {\bibinfo
  {journal} {Science}\ }\textbf {\bibinfo {volume} {331}},\ \bibinfo {pages}
  {189} (\bibinfo {year} {2011})}\BibitemShut {NoStop}%
\bibitem [{\citenamefont {{Hu}}\ \emph {et~al.}(2014)\citenamefont {{Hu}},
  \citenamefont {{Kaiser}}, \citenamefont {{Nicoletti}}, \citenamefont
  {{Hunt}}, \citenamefont {{Gierz}}, \citenamefont {{Hoffmann}}, \citenamefont
  {{Le Tacon}}, \citenamefont {{Loew}}, \citenamefont {{Keimer}},\ and\
  \citenamefont {{Cavalleri}}}]{Hu2014}%
  \BibitemOpen
  \bibfield  {author} {\bibinfo {author} {\bibfnamefont {W.}~\bibnamefont
  {{Hu}}}, \bibinfo {author} {\bibfnamefont {S.}~\bibnamefont {{Kaiser}}},
  \bibinfo {author} {\bibfnamefont {D.}~\bibnamefont {{Nicoletti}}}, \bibinfo
  {author} {\bibfnamefont {C.~R.}\ \bibnamefont {{Hunt}}}, \bibinfo {author}
  {\bibfnamefont {I.}~\bibnamefont {{Gierz}}}, \bibinfo {author} {\bibfnamefont
  {M.~C.}\ \bibnamefont {{Hoffmann}}}, \bibinfo {author} {\bibfnamefont
  {M.}~\bibnamefont {{Le Tacon}}}, \bibinfo {author} {\bibfnamefont
  {T.}~\bibnamefont {{Loew}}}, \bibinfo {author} {\bibfnamefont
  {B.}~\bibnamefont {{Keimer}}},\ and\ \bibinfo {author} {\bibfnamefont
  {A.}~\bibnamefont {{Cavalleri}}},\ }\bibfield  {title} {\bibinfo {title}
  {Optically enhanced coherent transport in {YBa$_{2}$Cu$_{3}$O$_{6.5}$} by
  ultrafast redistribution of interlayer coupling},\ }\href
  {https://doi.org/10.1038/nmat3963} {\bibfield  {journal} {\bibinfo  {journal}
  {Nat. Mater.}\ }\textbf {\bibinfo {volume} {13}},\ \bibinfo {pages} {705}
  (\bibinfo {year} {2014})}\BibitemShut {NoStop}%
\bibitem [{\citenamefont {Cremin}\ \emph {et~al.}(2019)\citenamefont {Cremin},
  \citenamefont {Zhang}, \citenamefont {Homes}, \citenamefont {Gu},
  \citenamefont {Sun}, \citenamefont {Fogler}, \citenamefont {Millis},
  \citenamefont {Basov},\ and\ \citenamefont {Averitt}}]{Cremin2019}%
  \BibitemOpen
  \bibfield  {author} {\bibinfo {author} {\bibfnamefont {K.~A.}\ \bibnamefont
  {Cremin}}, \bibinfo {author} {\bibfnamefont {J.}~\bibnamefont {Zhang}},
  \bibinfo {author} {\bibfnamefont {C.~C.}\ \bibnamefont {Homes}}, \bibinfo
  {author} {\bibfnamefont {G.~D.}\ \bibnamefont {Gu}}, \bibinfo {author}
  {\bibfnamefont {Z.}~\bibnamefont {Sun}}, \bibinfo {author} {\bibfnamefont
  {M.~M.}\ \bibnamefont {Fogler}}, \bibinfo {author} {\bibfnamefont {A.~J.}\
  \bibnamefont {Millis}}, \bibinfo {author} {\bibfnamefont {D.~N.}\
  \bibnamefont {Basov}},\ and\ \bibinfo {author} {\bibfnamefont {R.~D.}\
  \bibnamefont {Averitt}},\ }\bibfield  {title} {\bibinfo {title}
  {Photoenhanced metastable c-axis electrodynamics in stripe-ordered cuprate
  {La$_{1.885}$Ba$_{0.115}$CuO$_{4}$}},\ }\href
  {https://doi.org/10.1073/pnas.1908368116} {\bibfield  {journal} {\bibinfo
  {journal} {Proc. Natl. Acad. Sci. USA}\ }\textbf {\bibinfo {volume} {116}},\
  \bibinfo {pages} {19875} (\bibinfo {year} {2019})}\BibitemShut {NoStop}%
\bibitem [{\citenamefont {Kogar}\ \emph {et~al.}(2019)\citenamefont {Kogar},
  \citenamefont {Zong}, \citenamefont {Dolgirev}, \citenamefont {Shen},
  \citenamefont {Straquadine}, \citenamefont {Bie}, \citenamefont {Wang},
  \citenamefont {Rohwer}, \citenamefont {Tung}, \citenamefont {Yang},
  \citenamefont {Li}, \citenamefont {Yang}, \citenamefont {Weathersby},
  \citenamefont {Park}, \citenamefont {Kozina}, \citenamefont {Sie},
  \citenamefont {Wen}, \citenamefont {Jarillo-Herrero}, \citenamefont {Fisher},
  \citenamefont {Wang},\ and\ \citenamefont {Gedik}}]{Kogar2019}%
  \BibitemOpen
  \bibfield  {author} {\bibinfo {author} {\bibfnamefont {A.}~\bibnamefont
  {Kogar}}, \bibinfo {author} {\bibfnamefont {A.}~\bibnamefont {Zong}},
  \bibinfo {author} {\bibfnamefont {P.~E.}\ \bibnamefont {Dolgirev}}, \bibinfo
  {author} {\bibfnamefont {X.}~\bibnamefont {Shen}}, \bibinfo {author}
  {\bibfnamefont {J.}~\bibnamefont {Straquadine}}, \bibinfo {author}
  {\bibfnamefont {Y.-Q.}\ \bibnamefont {Bie}}, \bibinfo {author} {\bibfnamefont
  {X.}~\bibnamefont {Wang}}, \bibinfo {author} {\bibfnamefont {T.}~\bibnamefont
  {Rohwer}}, \bibinfo {author} {\bibfnamefont {I.-C.}\ \bibnamefont {Tung}},
  \bibinfo {author} {\bibfnamefont {Y.}~\bibnamefont {Yang}}, \bibinfo {author}
  {\bibfnamefont {R.}~\bibnamefont {Li}}, \bibinfo {author} {\bibfnamefont
  {J.}~\bibnamefont {Yang}}, \bibinfo {author} {\bibfnamefont {S.}~\bibnamefont
  {Weathersby}}, \bibinfo {author} {\bibfnamefont {S.}~\bibnamefont {Park}},
  \bibinfo {author} {\bibfnamefont {M.~E.}\ \bibnamefont {Kozina}}, \bibinfo
  {author} {\bibfnamefont {E.~J.}\ \bibnamefont {Sie}}, \bibinfo {author}
  {\bibfnamefont {H.}~\bibnamefont {Wen}}, \bibinfo {author} {\bibfnamefont
  {P.}~\bibnamefont {Jarillo-Herrero}}, \bibinfo {author} {\bibfnamefont
  {I.~R.}\ \bibnamefont {Fisher}}, \bibinfo {author} {\bibfnamefont
  {X.}~\bibnamefont {Wang}},\ and\ \bibinfo {author} {\bibfnamefont
  {N.}~\bibnamefont {Gedik}},\ }\bibfield  {title} {\bibinfo {title}
  {Light-induced charge density wave in {LaTe$_{3}$}},\ }\href
  {https://doi.org/10.1038/s41567-019-0705-3} {\bibfield  {journal} {\bibinfo
  {journal} {Nat. Phys.}\ }\textbf {\bibinfo {volume} {16}},\ \bibinfo {pages}
  {159} (\bibinfo {year} {2019})}\BibitemShut {NoStop}%
\bibitem [{\citenamefont {{Cosme}}\ \emph {et~al.}(2018)\citenamefont
  {{Cosme}}, \citenamefont {{Georges}}, \citenamefont {{Hemmerich}},\ and\
  \citenamefont {{Mathey}}}]{Cosme2018}%
  \BibitemOpen
  \bibfield  {author} {\bibinfo {author} {\bibfnamefont {J.~G.}\ \bibnamefont
  {{Cosme}}}, \bibinfo {author} {\bibfnamefont {C.}~\bibnamefont {{Georges}}},
  \bibinfo {author} {\bibfnamefont {A.}~\bibnamefont {{Hemmerich}}},\ and\
  \bibinfo {author} {\bibfnamefont {L.}~\bibnamefont {{Mathey}}},\ }\bibfield
  {title} {\bibinfo {title} {Dynamical control of order in a cavity-{BEC}
  system},\ }\href {https://doi.org/10.1103/PhysRevLett.121.153001} {\bibfield
  {journal} {\bibinfo  {journal} {Phys. Rev. Lett.}\ }\textbf {\bibinfo
  {volume} {121}},\ \bibinfo {pages} {153001} (\bibinfo {year}
  {2018})}\BibitemShut {NoStop}%
\bibitem [{\citenamefont {{Georges}}\ \emph {et~al.}(2018)\citenamefont
  {{Georges}}, \citenamefont {{Cosme}}, \citenamefont {{Mathey}},\ and\
  \citenamefont {{Hemmerich}}}]{Georges2018}%
  \BibitemOpen
  \bibfield  {author} {\bibinfo {author} {\bibfnamefont {C.}~\bibnamefont
  {{Georges}}}, \bibinfo {author} {\bibfnamefont {J.~G.}\ \bibnamefont
  {{Cosme}}}, \bibinfo {author} {\bibfnamefont {L.}~\bibnamefont {{Mathey}}},\
  and\ \bibinfo {author} {\bibfnamefont {A.}~\bibnamefont {{Hemmerich}}},\
  }\bibfield  {title} {\bibinfo {title} {Light-induced coherence in an
  atom-cavity system},\ }\href {https://doi.org/10.1103/PhysRevLett.121.220405}
  {\bibfield  {journal} {\bibinfo  {journal} {Phys. Rev. Lett.}\ }\textbf
  {\bibinfo {volume} {121}},\ \bibinfo {pages} {220405} (\bibinfo {year}
  {2018})}\BibitemShut {NoStop}%
\bibitem [{\citenamefont {Wilczek}(2013)}]{Wilczek2013}%
  \BibitemOpen
  \bibfield  {author} {\bibinfo {author} {\bibfnamefont {F.}~\bibnamefont
  {Wilczek}},\ }\bibfield  {title} {\bibinfo {title} {Superfluidity and
  space-time translation symmetry breaking},\ }\href
  {https://doi.org/10.1103/PhysRevLett.111.250402} {\bibfield  {journal}
  {\bibinfo  {journal} {Phys. Rev. Lett.}\ }\textbf {\bibinfo {volume} {111}},\
  \bibinfo {pages} {250402} (\bibinfo {year} {2013})}\BibitemShut {NoStop}%
\bibitem [{\citenamefont {{Sacha}}\ and\ \citenamefont
  {{Zakrzewski}}(2018)}]{Sacha2018}%
  \BibitemOpen
  \bibfield  {author} {\bibinfo {author} {\bibfnamefont {K.}~\bibnamefont
  {{Sacha}}}\ and\ \bibinfo {author} {\bibfnamefont {J.}~\bibnamefont
  {{Zakrzewski}}},\ }\bibfield  {title} {\bibinfo {title} {Time crystals: a
  review},\ }\href {https://doi.org/10.1088/1361-6633/aa8b38} {\bibfield
  {journal} {\bibinfo  {journal} {Rep. Prog. Phys.}\ }\textbf {\bibinfo
  {volume} {81}},\ \bibinfo {pages} {016401} (\bibinfo {year}
  {2018})}\BibitemShut {NoStop}%
\bibitem [{\citenamefont {{Gambetta}}\ \emph {et~al.}(2019)\citenamefont
  {{Gambetta}}, \citenamefont {{Carollo}}, \citenamefont {{Marcuzzi}},
  \citenamefont {{Garrahan}},\ and\ \citenamefont
  {{Lesanovsky}}}]{Gambetta2018}%
  \BibitemOpen
  \bibfield  {author} {\bibinfo {author} {\bibfnamefont {F.~M.}\ \bibnamefont
  {{Gambetta}}}, \bibinfo {author} {\bibfnamefont {F.}~\bibnamefont
  {{Carollo}}}, \bibinfo {author} {\bibfnamefont {M.}~\bibnamefont
  {{Marcuzzi}}}, \bibinfo {author} {\bibfnamefont {J.~P.}\ \bibnamefont
  {{Garrahan}}},\ and\ \bibinfo {author} {\bibfnamefont {I.}~\bibnamefont
  {{Lesanovsky}}},\ }\bibfield  {title} {\bibinfo {title} {Discrete time
  crystals in the absence of manifest symmetries or disorder in open quantum
  systems},\ }\href {https://doi.org/10.1103/PhysRevLett.122.015701} {\bibfield
   {journal} {\bibinfo  {journal} {Phys. Rev. Lett.}\ }\textbf {\bibinfo
  {volume} {122}},\ \bibinfo {pages} {015701} (\bibinfo {year}
  {2019})}\BibitemShut {NoStop}%
\bibitem [{\citenamefont {{Buca}}\ \emph {et~al.}(2019)\citenamefont {{Buca}},
  \citenamefont {{Tindall}},\ and\ \citenamefont {{Jaksch}}}]{Buca2019}%
  \BibitemOpen
  \bibfield  {author} {\bibinfo {author} {\bibfnamefont {B.}~\bibnamefont
  {{Buca}}}, \bibinfo {author} {\bibfnamefont {J.}~\bibnamefont {{Tindall}}},\
  and\ \bibinfo {author} {\bibfnamefont {D.}~\bibnamefont {{Jaksch}}},\
  }\bibfield  {title} {\bibinfo {title} {Non-stationary coherent quantum
  many-body dynamics through dissipation},\ }\href
  {https://doi.org/10.1038/s41467-019-09757-y} {\bibfield  {journal} {\bibinfo
  {journal} {Nat. Commun.}\ }\textbf {\bibinfo {volume} {10}},\ \bibinfo
  {pages} {1730} (\bibinfo {year} {2019})}\BibitemShut {NoStop}%
\bibitem [{\citenamefont {Heugel}\ \emph {et~al.}(2019)\citenamefont {Heugel},
  \citenamefont {Oscity}, \citenamefont {Eichler}, \citenamefont {Zilberberg},\
  and\ \citenamefont {Chitra}}]{Heugel2019}%
  \BibitemOpen
  \bibfield  {author} {\bibinfo {author} {\bibfnamefont {T.~L.}\ \bibnamefont
  {Heugel}}, \bibinfo {author} {\bibfnamefont {M.}~\bibnamefont {Oscity}},
  \bibinfo {author} {\bibfnamefont {A.}~\bibnamefont {Eichler}}, \bibinfo
  {author} {\bibfnamefont {O.}~\bibnamefont {Zilberberg}},\ and\ \bibinfo
  {author} {\bibfnamefont {R.}~\bibnamefont {Chitra}},\ }\bibfield  {title}
  {\bibinfo {title} {Classical many-body time crystals},\ }\href
  {https://doi.org/10.1103/PhysRevLett.123.124301} {\bibfield  {journal}
  {\bibinfo  {journal} {Phys. Rev. Lett.}\ }\textbf {\bibinfo {volume} {123}},\
  \bibinfo {pages} {124301} (\bibinfo {year} {2019})}\BibitemShut {NoStop}%
\bibitem [{\citenamefont {{Else}}\ \emph {et~al.}(2020)\citenamefont {{Else}},
  \citenamefont {{Monroe}}, \citenamefont {{Nayak}},\ and\ \citenamefont
  {{Yao}}}]{Else2020}%
  \BibitemOpen
  \bibfield  {author} {\bibinfo {author} {\bibfnamefont {D.~V.}\ \bibnamefont
  {{Else}}}, \bibinfo {author} {\bibfnamefont {C.}~\bibnamefont {{Monroe}}},
  \bibinfo {author} {\bibfnamefont {C.}~\bibnamefont {{Nayak}}},\ and\ \bibinfo
  {author} {\bibfnamefont {N.~Y.}\ \bibnamefont {{Yao}}},\ }\bibfield  {title}
  {\bibinfo {title} {Discrete time crystals},\ }\href
  {https://doi.org/10.1146/annurev-conmatphys-031119-050658} {\bibfield
  {journal} {\bibinfo  {journal} {{Annu. Rev. Condens. Matter Phys.}}\ }\textbf
  {\bibinfo {volume} {11}},\ \bibinfo {pages} {467} (\bibinfo {year}
  {2020})}\BibitemShut {NoStop}%
\bibitem [{\citenamefont {Yao}\ \emph {et~al.}(2020)\citenamefont {Yao},
  \citenamefont {Nayak}, \citenamefont {Balents},\ and\ \citenamefont
  {Zaletel}}]{Yao2020}%
  \BibitemOpen
  \bibfield  {author} {\bibinfo {author} {\bibfnamefont {N.~Y.}\ \bibnamefont
  {Yao}}, \bibinfo {author} {\bibfnamefont {C.}~\bibnamefont {Nayak}}, \bibinfo
  {author} {\bibfnamefont {L.}~\bibnamefont {Balents}},\ and\ \bibinfo {author}
  {\bibfnamefont {M.~P.}\ \bibnamefont {Zaletel}},\ }\bibfield  {title}
  {\bibinfo {title} {Classical discrete time crystals},\ }\href
  {https://doi.org/10.1038/s41567-019-0782-3} {\bibfield  {journal} {\bibinfo
  {journal} {Nat. Phys.}\ }\textbf {\bibinfo {volume} {16}},\ \bibinfo {pages}
  {438} (\bibinfo {year} {2020})}\BibitemShut {NoStop}%
\bibitem [{\citenamefont {{Choi}}\ \emph {et~al.}(2017)\citenamefont {{Choi}},
  \citenamefont {{Choi}}, \citenamefont {{Landig}}, \citenamefont {{Kucsko}},
  \citenamefont {{Zhou}}, \citenamefont {{Isoya}}, \citenamefont {{Jelezko}},
  \citenamefont {{Onoda}}, \citenamefont {{Sumiya}}, \citenamefont {{Khemani}},
  \citenamefont {{von Keyserlingk}}, \citenamefont {{Yao}}, \citenamefont
  {{Demler}},\ and\ \citenamefont {{Lukin}}}]{Choi2017}%
  \BibitemOpen
  \bibfield  {author} {\bibinfo {author} {\bibfnamefont {S.}~\bibnamefont
  {{Choi}}}, \bibinfo {author} {\bibfnamefont {J.}~\bibnamefont {{Choi}}},
  \bibinfo {author} {\bibfnamefont {R.}~\bibnamefont {{Landig}}}, \bibinfo
  {author} {\bibfnamefont {G.}~\bibnamefont {{Kucsko}}}, \bibinfo {author}
  {\bibfnamefont {H.}~\bibnamefont {{Zhou}}}, \bibinfo {author} {\bibfnamefont
  {J.}~\bibnamefont {{Isoya}}}, \bibinfo {author} {\bibfnamefont
  {F.}~\bibnamefont {{Jelezko}}}, \bibinfo {author} {\bibfnamefont
  {S.}~\bibnamefont {{Onoda}}}, \bibinfo {author} {\bibfnamefont
  {H.}~\bibnamefont {{Sumiya}}}, \bibinfo {author} {\bibfnamefont
  {V.}~\bibnamefont {{Khemani}}}, \bibinfo {author} {\bibfnamefont
  {C.}~\bibnamefont {{von Keyserlingk}}}, \bibinfo {author} {\bibfnamefont
  {N.~Y.}\ \bibnamefont {{Yao}}}, \bibinfo {author} {\bibfnamefont
  {E.}~\bibnamefont {{Demler}}},\ and\ \bibinfo {author} {\bibfnamefont
  {M.~D.}\ \bibnamefont {{Lukin}}},\ }\bibfield  {title} {\bibinfo {title}
  {Observation of discrete time-crystalline order in a disordered dipolar
  many-body system},\ }\href {https://doi.org/10.1038/nature21426} {\bibfield
  {journal} {\bibinfo  {journal} {Nature}\ }\textbf {\bibinfo {volume} {543}},\
  \bibinfo {pages} {221} (\bibinfo {year} {2017})}\BibitemShut {NoStop}%
\bibitem [{\citenamefont {{Zhang}}\ \emph {et~al.}(2017)\citenamefont
  {{Zhang}}, \citenamefont {{Hess}}, \citenamefont {{Kyprianidis}},
  \citenamefont {{Becker}}, \citenamefont {{Lee}}, \citenamefont {{Smith}},
  \citenamefont {{Pagano}}, \citenamefont {{Potirniche}}, \citenamefont
  {{Potter}}, \citenamefont {{Vishwanath}}, \citenamefont {{Yao}},\ and\
  \citenamefont {{Monroe}}}]{Zhang2017}%
  \BibitemOpen
  \bibfield  {author} {\bibinfo {author} {\bibfnamefont {J.}~\bibnamefont
  {{Zhang}}}, \bibinfo {author} {\bibfnamefont {P.~W.}\ \bibnamefont {{Hess}}},
  \bibinfo {author} {\bibfnamefont {A.}~\bibnamefont {{Kyprianidis}}}, \bibinfo
  {author} {\bibfnamefont {P.}~\bibnamefont {{Becker}}}, \bibinfo {author}
  {\bibfnamefont {A.}~\bibnamefont {{Lee}}}, \bibinfo {author} {\bibfnamefont
  {J.}~\bibnamefont {{Smith}}}, \bibinfo {author} {\bibfnamefont
  {G.}~\bibnamefont {{Pagano}}}, \bibinfo {author} {\bibfnamefont {I.-D.}\
  \bibnamefont {{Potirniche}}}, \bibinfo {author} {\bibfnamefont {A.~C.}\
  \bibnamefont {{Potter}}}, \bibinfo {author} {\bibfnamefont {A.}~\bibnamefont
  {{Vishwanath}}}, \bibinfo {author} {\bibfnamefont {N.~Y.}\ \bibnamefont
  {{Yao}}},\ and\ \bibinfo {author} {\bibfnamefont {C.}~\bibnamefont
  {{Monroe}}},\ }\bibfield  {title} {\bibinfo {title} {Observation of a
  discrete time crystal},\ }\href {https://doi.org/10.1038/nature21413}
  {\bibfield  {journal} {\bibinfo  {journal} {Nature}\ }\textbf {\bibinfo
  {volume} {543}},\ \bibinfo {pages} {217} (\bibinfo {year}
  {2017})}\BibitemShut {NoStop}%
\bibitem [{\citenamefont {Chew}\ \emph {et~al.}(2020)\citenamefont {Chew},
  \citenamefont {Mross},\ and\ \citenamefont {Alicea}}]{Chew2020}%
  \BibitemOpen
  \bibfield  {author} {\bibinfo {author} {\bibfnamefont {A.}~\bibnamefont
  {Chew}}, \bibinfo {author} {\bibfnamefont {D.~F.}\ \bibnamefont {Mross}},\
  and\ \bibinfo {author} {\bibfnamefont {J.}~\bibnamefont {Alicea}},\
  }\bibfield  {title} {\bibinfo {title} {Time-crystalline topological
  superconductors},\ }\href {https://doi.org/10.1103/PhysRevLett.124.096802}
  {\bibfield  {journal} {\bibinfo  {journal} {Phys. Rev. Lett.}\ }\textbf
  {\bibinfo {volume} {124}},\ \bibinfo {pages} {096802} (\bibinfo {year}
  {2020})}\BibitemShut {NoStop}%
\bibitem [{\citenamefont {Varma}(2002)}]{Varma2002}%
  \BibitemOpen
  \bibfield  {author} {\bibinfo {author} {\bibfnamefont {C.~M.}\ \bibnamefont
  {Varma}},\ }\bibfield  {title} {\bibinfo {title} {Higgs boson in
  superconductors},\ }\href {https://doi.org/10.1023/A:1013890507658}
  {\bibfield  {journal} {\bibinfo  {journal} {J. Low Temp. Phys.}\ }\textbf
  {\bibinfo {volume} {126}},\ \bibinfo {pages} {901} (\bibinfo {year}
  {2002})}\BibitemShut {NoStop}%
\bibitem [{\citenamefont {Pekker}\ and\ \citenamefont
  {Varma}(2015)}]{Pekker2015}%
  \BibitemOpen
  \bibfield  {author} {\bibinfo {author} {\bibfnamefont {D.}~\bibnamefont
  {Pekker}}\ and\ \bibinfo {author} {\bibfnamefont {C.}~\bibnamefont {Varma}},\
  }\bibfield  {title} {\bibinfo {title} {{Amplitude/Higgs} modes in condensed
  matter physics},\ }\href
  {https://doi.org/10.1146/annurev-conmatphys-031214-014350} {\bibfield
  {journal} {\bibinfo  {journal} {Annu. Rev. Condens. Matter Phys.}\ }\textbf
  {\bibinfo {volume} {6}},\ \bibinfo {pages} {269} (\bibinfo {year}
  {2015})}\BibitemShut {NoStop}%
\bibitem [{\citenamefont {Matsunaga}\ \emph {et~al.}(2014)\citenamefont
  {Matsunaga}, \citenamefont {Tsuji}, \citenamefont {Fujita}, \citenamefont
  {Sugioka}, \citenamefont {Makise}, \citenamefont {Uzawa}, \citenamefont
  {Terai}, \citenamefont {Wang}, \citenamefont {Aoki},\ and\ \citenamefont
  {Shimano}}]{Matsunaga2014}%
  \BibitemOpen
  \bibfield  {author} {\bibinfo {author} {\bibfnamefont {R.}~\bibnamefont
  {Matsunaga}}, \bibinfo {author} {\bibfnamefont {N.}~\bibnamefont {Tsuji}},
  \bibinfo {author} {\bibfnamefont {H.}~\bibnamefont {Fujita}}, \bibinfo
  {author} {\bibfnamefont {A.}~\bibnamefont {Sugioka}}, \bibinfo {author}
  {\bibfnamefont {K.}~\bibnamefont {Makise}}, \bibinfo {author} {\bibfnamefont
  {Y.}~\bibnamefont {Uzawa}}, \bibinfo {author} {\bibfnamefont
  {H.}~\bibnamefont {Terai}}, \bibinfo {author} {\bibfnamefont
  {Z.}~\bibnamefont {Wang}}, \bibinfo {author} {\bibfnamefont {H.}~\bibnamefont
  {Aoki}},\ and\ \bibinfo {author} {\bibfnamefont {R.}~\bibnamefont
  {Shimano}},\ }\bibfield  {title} {\bibinfo {title} {Light-induced collective
  pseudospin precession resonating with {Higgs} mode in a superconductor},\
  }\href {https://doi.org/10.1126/science.1254697} {\bibfield  {journal}
  {\bibinfo  {journal} {Science}\ }\textbf {\bibinfo {volume} {345}},\ \bibinfo
  {pages} {1145} (\bibinfo {year} {2014})}\BibitemShut {NoStop}%
\bibitem [{\citenamefont {Tsuji}\ and\ \citenamefont {Aoki}(2015)}]{Tsuji2015}%
  \BibitemOpen
  \bibfield  {author} {\bibinfo {author} {\bibfnamefont {N.}~\bibnamefont
  {Tsuji}}\ and\ \bibinfo {author} {\bibfnamefont {H.}~\bibnamefont {Aoki}},\
  }\bibfield  {title} {\bibinfo {title} {Theory of {Anderson} pseudospin
  resonance with {Higgs} mode in superconductors},\ }\href
  {https://doi.org/10.1103/PhysRevB.92.064508} {\bibfield  {journal} {\bibinfo
  {journal} {Phys. Rev. B}\ }\textbf {\bibinfo {volume} {92}},\ \bibinfo
  {pages} {064508} (\bibinfo {year} {2015})}\BibitemShut {NoStop}%
\bibitem [{\citenamefont {Nakamura}\ \emph {et~al.}(2019)\citenamefont
  {Nakamura}, \citenamefont {Iida}, \citenamefont {Murotani}, \citenamefont
  {Matsunaga}, \citenamefont {Terai},\ and\ \citenamefont
  {Shimano}}]{Nakamura2019}%
  \BibitemOpen
  \bibfield  {author} {\bibinfo {author} {\bibfnamefont {S.}~\bibnamefont
  {Nakamura}}, \bibinfo {author} {\bibfnamefont {Y.}~\bibnamefont {Iida}},
  \bibinfo {author} {\bibfnamefont {Y.}~\bibnamefont {Murotani}}, \bibinfo
  {author} {\bibfnamefont {R.}~\bibnamefont {Matsunaga}}, \bibinfo {author}
  {\bibfnamefont {H.}~\bibnamefont {Terai}},\ and\ \bibinfo {author}
  {\bibfnamefont {R.}~\bibnamefont {Shimano}},\ }\bibfield  {title} {\bibinfo
  {title} {Infrared activation of the {Higgs} mode by supercurrent injection in
  superconducting {NbN}},\ }\href
  {https://doi.org/10.1103/PhysRevLett.122.257001} {\bibfield  {journal}
  {\bibinfo  {journal} {Phys. Rev. Lett.}\ }\textbf {\bibinfo {volume} {122}},\
  \bibinfo {pages} {257001} (\bibinfo {year} {2019})}\BibitemShut {NoStop}%
\bibitem [{\citenamefont {Shimano}\ and\ \citenamefont
  {Tsuji}(2020)}]{Shimano2020}%
  \BibitemOpen
  \bibfield  {author} {\bibinfo {author} {\bibfnamefont {R.}~\bibnamefont
  {Shimano}}\ and\ \bibinfo {author} {\bibfnamefont {N.}~\bibnamefont
  {Tsuji}},\ }\bibfield  {title} {\bibinfo {title} {Higgs mode in
  superconductors},\ }\href
  {https://doi.org/10.1146/annurev-conmatphys-031119-050813} {\bibfield
  {journal} {\bibinfo  {journal} {Annu. Rev. Condens. Matter Phys.}\ }\textbf
  {\bibinfo {volume} {11}},\ \bibinfo {pages} {103} (\bibinfo {year}
  {2020})}\BibitemShut {NoStop}%
\bibitem [{\citenamefont {Peronaci}\ \emph {et~al.}(2015)\citenamefont
  {Peronaci}, \citenamefont {Schir\'o},\ and\ \citenamefont
  {Capone}}]{Peronaci2015}%
  \BibitemOpen
  \bibfield  {author} {\bibinfo {author} {\bibfnamefont {F.}~\bibnamefont
  {Peronaci}}, \bibinfo {author} {\bibfnamefont {M.}~\bibnamefont {Schir\'o}},\
  and\ \bibinfo {author} {\bibfnamefont {M.}~\bibnamefont {Capone}},\
  }\bibfield  {title} {\bibinfo {title} {Transient dynamics of $d$-wave
  superconductors after a sudden excitation},\ }\href
  {https://doi.org/10.1103/PhysRevLett.115.257001} {\bibfield  {journal}
  {\bibinfo  {journal} {Phys. Rev. Lett.}\ }\textbf {\bibinfo {volume} {115}},\
  \bibinfo {pages} {257001} (\bibinfo {year} {2015})}\BibitemShut {NoStop}%
\bibitem [{\citenamefont {Katsumi}\ \emph {et~al.}(2018)\citenamefont
  {Katsumi}, \citenamefont {Tsuji}, \citenamefont {Hamada}, \citenamefont
  {Matsunaga}, \citenamefont {Schneeloch}, \citenamefont {Zhong}, \citenamefont
  {Gu}, \citenamefont {Aoki}, \citenamefont {Gallais},\ and\ \citenamefont
  {Shimano}}]{Katsumi2018}%
  \BibitemOpen
  \bibfield  {author} {\bibinfo {author} {\bibfnamefont {K.}~\bibnamefont
  {Katsumi}}, \bibinfo {author} {\bibfnamefont {N.}~\bibnamefont {Tsuji}},
  \bibinfo {author} {\bibfnamefont {Y.~I.}\ \bibnamefont {Hamada}}, \bibinfo
  {author} {\bibfnamefont {R.}~\bibnamefont {Matsunaga}}, \bibinfo {author}
  {\bibfnamefont {J.}~\bibnamefont {Schneeloch}}, \bibinfo {author}
  {\bibfnamefont {R.~D.}\ \bibnamefont {Zhong}}, \bibinfo {author}
  {\bibfnamefont {G.~D.}\ \bibnamefont {Gu}}, \bibinfo {author} {\bibfnamefont
  {H.}~\bibnamefont {Aoki}}, \bibinfo {author} {\bibfnamefont {Y.}~\bibnamefont
  {Gallais}},\ and\ \bibinfo {author} {\bibfnamefont {R.}~\bibnamefont
  {Shimano}},\ }\bibfield  {title} {\bibinfo {title} {Higgs mode in the
  $d$-wave superconductor
  {${\mathrm{Bi}}_{2}{\mathrm{Sr}}_{2}{\mathrm{CaCu}}_{2}{\mathrm{O}}_{8+x}$}
  driven by an intense terahertz pulse},\ }\href
  {https://doi.org/10.1103/PhysRevLett.120.117001} {\bibfield  {journal}
  {\bibinfo  {journal} {Phys. Rev. Lett.}\ }\textbf {\bibinfo {volume} {120}},\
  \bibinfo {pages} {117001} (\bibinfo {year} {2018})}\BibitemShut {NoStop}%
\bibitem [{\citenamefont {Buzzi}\ \emph {et~al.}(2019)\citenamefont {Buzzi},
  \citenamefont {Jotzu}, \citenamefont {Cavalleri}, \citenamefont {Cirac},
  \citenamefont {Demler}, \citenamefont {Halperin}, \citenamefont {Lukin},
  \citenamefont {Shi}, \citenamefont {Wang},\ and\ \citenamefont
  {Podolsky}}]{Buzzi2019}%
  \BibitemOpen
  \bibfield  {author} {\bibinfo {author} {\bibfnamefont {M.}~\bibnamefont
  {Buzzi}}, \bibinfo {author} {\bibfnamefont {G.}~\bibnamefont {Jotzu}},
  \bibinfo {author} {\bibfnamefont {A.}~\bibnamefont {Cavalleri}}, \bibinfo
  {author} {\bibfnamefont {J.~I.}\ \bibnamefont {Cirac}}, \bibinfo {author}
  {\bibfnamefont {E.~A.}\ \bibnamefont {Demler}}, \bibinfo {author}
  {\bibfnamefont {B.~I.}\ \bibnamefont {Halperin}}, \bibinfo {author}
  {\bibfnamefont {M.~D.}\ \bibnamefont {Lukin}}, \bibinfo {author}
  {\bibfnamefont {T.}~\bibnamefont {Shi}}, \bibinfo {author} {\bibfnamefont
  {Y.}~\bibnamefont {Wang}},\ and\ \bibinfo {author} {\bibfnamefont
  {D.}~\bibnamefont {Podolsky}},\ }\bibfield  {title} {\bibinfo {title}
  {Higgs-mediated optical amplification in a non-equilibrium superconductor},\
  }\href@noop {} {\bibfield  {journal} {\bibinfo  {journal} {arXiv e-prints}\ }
  (\bibinfo {year} {2019})},\ \Eprint {https://arxiv.org/abs/1908.10879}
  {arXiv:1908.10879 [cond-mat.supr-con]} \BibitemShut {NoStop}%
\bibitem [{\citenamefont {Chu}\ \emph {et~al.}(2020)\citenamefont {Chu},
  \citenamefont {Kim}, \citenamefont {Katsumi}, \citenamefont {Kovalev},
  \citenamefont {Dawson}, \citenamefont {Schwarz}, \citenamefont {Yoshikawa},
  \citenamefont {Kim}, \citenamefont {Putzky}, \citenamefont {Li},
  \citenamefont {Raffy}, \citenamefont {Germanskiy}, \citenamefont {Deinert},
  \citenamefont {Awari}, \citenamefont {Ilyakov}, \citenamefont {Green},
  \citenamefont {Chen}, \citenamefont {Bawatna}, \citenamefont {Christiani},
  \citenamefont {Logvenov}, \citenamefont {Gallais}, \citenamefont {Boris},
  \citenamefont {Keimer}, \citenamefont {Schnyder}, \citenamefont {Manske},
  \citenamefont {Gensch}, \citenamefont {Wang}, \citenamefont {Shimano},\ and\
  \citenamefont {Kaiser}}]{Chu2020}%
  \BibitemOpen
  \bibfield  {author} {\bibinfo {author} {\bibfnamefont {H.}~\bibnamefont
  {Chu}}, \bibinfo {author} {\bibfnamefont {M.-J.}\ \bibnamefont {Kim}},
  \bibinfo {author} {\bibfnamefont {K.}~\bibnamefont {Katsumi}}, \bibinfo
  {author} {\bibfnamefont {S.}~\bibnamefont {Kovalev}}, \bibinfo {author}
  {\bibfnamefont {R.~D.}\ \bibnamefont {Dawson}}, \bibinfo {author}
  {\bibfnamefont {L.}~\bibnamefont {Schwarz}}, \bibinfo {author} {\bibfnamefont
  {N.}~\bibnamefont {Yoshikawa}}, \bibinfo {author} {\bibfnamefont
  {G.}~\bibnamefont {Kim}}, \bibinfo {author} {\bibfnamefont {D.}~\bibnamefont
  {Putzky}}, \bibinfo {author} {\bibfnamefont {Z.~Z.}\ \bibnamefont {Li}},
  \bibinfo {author} {\bibfnamefont {H.}~\bibnamefont {Raffy}}, \bibinfo
  {author} {\bibfnamefont {S.}~\bibnamefont {Germanskiy}}, \bibinfo {author}
  {\bibfnamefont {J.-C.}\ \bibnamefont {Deinert}}, \bibinfo {author}
  {\bibfnamefont {N.}~\bibnamefont {Awari}}, \bibinfo {author} {\bibfnamefont
  {I.}~\bibnamefont {Ilyakov}}, \bibinfo {author} {\bibfnamefont
  {B.}~\bibnamefont {Green}}, \bibinfo {author} {\bibfnamefont
  {M.}~\bibnamefont {Chen}}, \bibinfo {author} {\bibfnamefont {M.}~\bibnamefont
  {Bawatna}}, \bibinfo {author} {\bibfnamefont {G.}~\bibnamefont {Christiani}},
  \bibinfo {author} {\bibfnamefont {G.}~\bibnamefont {Logvenov}}, \bibinfo
  {author} {\bibfnamefont {Y.}~\bibnamefont {Gallais}}, \bibinfo {author}
  {\bibfnamefont {A.~V.}\ \bibnamefont {Boris}}, \bibinfo {author}
  {\bibfnamefont {B.}~\bibnamefont {Keimer}}, \bibinfo {author} {\bibfnamefont
  {A.~P.}\ \bibnamefont {Schnyder}}, \bibinfo {author} {\bibfnamefont
  {D.}~\bibnamefont {Manske}}, \bibinfo {author} {\bibfnamefont
  {M.}~\bibnamefont {Gensch}}, \bibinfo {author} {\bibfnamefont
  {Z.}~\bibnamefont {Wang}}, \bibinfo {author} {\bibfnamefont {R.}~\bibnamefont
  {Shimano}},\ and\ \bibinfo {author} {\bibfnamefont {S.}~\bibnamefont
  {Kaiser}},\ }\bibfield  {title} {\bibinfo {title} {Phase-resolved {Higgs}
  response in superconducting cuprates},\ }\href
  {https://doi.org/10.1038/s41467-020-15613-1} {\bibfield  {journal} {\bibinfo
  {journal} {Nat. Commun.}\ }\textbf {\bibinfo {volume} {11}},\ \bibinfo
  {pages} {1793} (\bibinfo {year} {2020})}\BibitemShut {NoStop}%
\bibitem [{\citenamefont {Schwarz}\ \emph {et~al.}(2020)\citenamefont
  {Schwarz}, \citenamefont {Fauseweh}, \citenamefont {Tsuji}, \citenamefont
  {Cheng}, \citenamefont {Bittner}, \citenamefont {Krull}, \citenamefont
  {Berciu}, \citenamefont {Uhrig}, \citenamefont {Schnyder}, \citenamefont
  {Kaiser},\ and\ \citenamefont {Manske}}]{Schwarz2020}%
  \BibitemOpen
  \bibfield  {author} {\bibinfo {author} {\bibfnamefont {L.}~\bibnamefont
  {Schwarz}}, \bibinfo {author} {\bibfnamefont {B.}~\bibnamefont {Fauseweh}},
  \bibinfo {author} {\bibfnamefont {N.}~\bibnamefont {Tsuji}}, \bibinfo
  {author} {\bibfnamefont {N.}~\bibnamefont {Cheng}}, \bibinfo {author}
  {\bibfnamefont {N.}~\bibnamefont {Bittner}}, \bibinfo {author} {\bibfnamefont
  {H.}~\bibnamefont {Krull}}, \bibinfo {author} {\bibfnamefont
  {M.}~\bibnamefont {Berciu}}, \bibinfo {author} {\bibfnamefont {G.~S.}\
  \bibnamefont {Uhrig}}, \bibinfo {author} {\bibfnamefont {A.~P.}\ \bibnamefont
  {Schnyder}}, \bibinfo {author} {\bibfnamefont {S.}~\bibnamefont {Kaiser}},\
  and\ \bibinfo {author} {\bibfnamefont {D.}~\bibnamefont {Manske}},\
  }\bibfield  {title} {\bibinfo {title} {Classification and characterization of
  nonequilibrium {Higgs} modes in unconventional superconductors},\ }\href
  {https://doi.org/10.1038/s41467-019-13763-5} {\bibfield  {journal} {\bibinfo
  {journal} {Nat. Commun.}\ }\textbf {\bibinfo {volume} {11}},\ \bibinfo
  {pages} {287} (\bibinfo {year} {2020})}\BibitemShut {NoStop}%
\bibitem [{\citenamefont {Puviani}\ \emph {et~al.}(2020)\citenamefont
  {Puviani}, \citenamefont {Schwarz}, \citenamefont {Zhang}, \citenamefont
  {Kaiser},\ and\ \citenamefont {Manske}}]{Puviani2020}%
  \BibitemOpen
  \bibfield  {author} {\bibinfo {author} {\bibfnamefont {M.}~\bibnamefont
  {Puviani}}, \bibinfo {author} {\bibfnamefont {L.}~\bibnamefont {Schwarz}},
  \bibinfo {author} {\bibfnamefont {X.-X.}\ \bibnamefont {Zhang}}, \bibinfo
  {author} {\bibfnamefont {S.}~\bibnamefont {Kaiser}},\ and\ \bibinfo {author}
  {\bibfnamefont {D.}~\bibnamefont {Manske}},\ }\bibfield  {title} {\bibinfo
  {title} {Current-assisted {Raman} activation of the {Higgs} mode in
  superconductors},\ }\href {https://doi.org/10.1103/PhysRevB.101.220507}
  {\bibfield  {journal} {\bibinfo  {journal} {Phys. Rev. B}\ }\textbf {\bibinfo
  {volume} {101}},\ \bibinfo {pages} {220507} (\bibinfo {year}
  {2020})}\BibitemShut {NoStop}%
\bibitem [{\citenamefont {Yang}\ and\ \citenamefont {Wu}(2020)}]{Yang2020}%
  \BibitemOpen
  \bibfield  {author} {\bibinfo {author} {\bibfnamefont {F.}~\bibnamefont
  {Yang}}\ and\ \bibinfo {author} {\bibfnamefont {M.~W.}\ \bibnamefont {Wu}},\
  }\bibfield  {title} {\bibinfo {title} {Theory of {Higgs} modes in $d$-wave
  superconductors},\ }\href {https://doi.org/10.1103/PhysRevB.102.014511}
  {\bibfield  {journal} {\bibinfo  {journal} {Phys. Rev. B}\ }\textbf {\bibinfo
  {volume} {102}},\ \bibinfo {pages} {014511} (\bibinfo {year}
  {2020})}\BibitemShut {NoStop}%
\bibitem [{\citenamefont {Yuzbashyan}\ \emph {et~al.}(2006)\citenamefont
  {Yuzbashyan}, \citenamefont {Tsyplyatyev},\ and\ \citenamefont
  {Altshuler}}]{Yuzbashyan2006}%
  \BibitemOpen
  \bibfield  {author} {\bibinfo {author} {\bibfnamefont {E.~A.}\ \bibnamefont
  {Yuzbashyan}}, \bibinfo {author} {\bibfnamefont {O.}~\bibnamefont
  {Tsyplyatyev}},\ and\ \bibinfo {author} {\bibfnamefont {B.~L.}\ \bibnamefont
  {Altshuler}},\ }\bibfield  {title} {\bibinfo {title} {Relaxation and
  persistent oscillations of the order parameter in fermionic condensates},\
  }\href {https://doi.org/10.1103/PhysRevLett.96.097005} {\bibfield  {journal}
  {\bibinfo  {journal} {Phys. Rev. Lett.}\ }\textbf {\bibinfo {volume} {96}},\
  \bibinfo {pages} {097005} (\bibinfo {year} {2006})}\BibitemShut {NoStop}%
\bibitem [{\citenamefont {Koyama}\ and\ \citenamefont
  {Tachiki}(1996)}]{Koyama1996}%
  \BibitemOpen
  \bibfield  {author} {\bibinfo {author} {\bibfnamefont {T.}~\bibnamefont
  {Koyama}}\ and\ \bibinfo {author} {\bibfnamefont {M.}~\bibnamefont
  {Tachiki}},\ }\bibfield  {title} {\bibinfo {title} {{I-V} characteristics of
  {Josephson}-coupled layered superconductors with longitudinal plasma
  excitations},\ }\href {https://doi.org/10.1103/PhysRevB.54.16183} {\bibfield
  {journal} {\bibinfo  {journal} {Phys. Rev. B}\ }\textbf {\bibinfo {volume}
  {54}},\ \bibinfo {pages} {16183} (\bibinfo {year} {1996})}\BibitemShut
  {NoStop}%
\bibitem [{\citenamefont {van~der Marel}\ and\ \citenamefont
  {Tsvetkov}(2001)}]{Marel2001}%
  \BibitemOpen
  \bibfield  {author} {\bibinfo {author} {\bibfnamefont {D.}~\bibnamefont
  {van~der Marel}}\ and\ \bibinfo {author} {\bibfnamefont {A.~A.}\ \bibnamefont
  {Tsvetkov}},\ }\bibfield  {title} {\bibinfo {title} {Transverse-optical
  {Josephson} plasmons: Equations of motion},\ }\href
  {https://doi.org/10.1103/PhysRevB.64.024530} {\bibfield  {journal} {\bibinfo
  {journal} {Phys. Rev. B}\ }\textbf {\bibinfo {volume} {64}},\ \bibinfo
  {pages} {024530} (\bibinfo {year} {2001})}\BibitemShut {NoStop}%
\bibitem [{\citenamefont {{Koyama}}(2002)}]{Koyama2002}%
  \BibitemOpen
  \bibfield  {author} {\bibinfo {author} {\bibfnamefont {T.}~\bibnamefont
  {{Koyama}}},\ }\bibfield  {title} {\bibinfo {title} {Josephson plasma
  resonances and optical properties in high-${T}_{c}$ superconductors with
  alternating junction parameters},\ }\href
  {https://doi.org/10.1143/JPSJ.71.2986} {\bibfield  {journal} {\bibinfo
  {journal} {J. Phys. Soc. Jpn.}\ }\textbf {\bibinfo {volume} {71}},\ \bibinfo
  {pages} {2986} (\bibinfo {year} {2002})}\BibitemShut {NoStop}%
\bibitem [{\citenamefont {Shibata}\ and\ \citenamefont
  {Yamada}(1998)}]{Shibata1998}%
  \BibitemOpen
  \bibfield  {author} {\bibinfo {author} {\bibfnamefont {H.}~\bibnamefont
  {Shibata}}\ and\ \bibinfo {author} {\bibfnamefont {T.}~\bibnamefont
  {Yamada}},\ }\bibfield  {title} {\bibinfo {title} {Double josephson plasma
  resonance in {$T^{*}$} phase
  {${\mathrm{SmLa}}_{1\ensuremath{-}\mathit{x}}{\mathrm{Sr}}_{\mathit{x}}{\mathrm{CuO}}_{4\ensuremath{-}\mathit{\ensuremath{\delta}}}$}},\
  }\href {https://doi.org/10.1103/PhysRevLett.81.3519} {\bibfield  {journal}
  {\bibinfo  {journal} {Phys. Rev. Lett.}\ }\textbf {\bibinfo {volume} {81}},\
  \bibinfo {pages} {3519} (\bibinfo {year} {1998})}\BibitemShut {NoStop}%
\bibitem [{\citenamefont {Duli\ifmmode~\acute{c}\else \'{c}\fi{}}\ \emph
  {et~al.}(2001)\citenamefont {Duli\ifmmode~\acute{c}\else \'{c}\fi{}},
  \citenamefont {Pimenov}, \citenamefont {van~der Marel}, \citenamefont
  {Broun}, \citenamefont {Kamal}, \citenamefont {Hardy}, \citenamefont
  {Tsvetkov}, \citenamefont {Sutjaha}, \citenamefont {Liang}, \citenamefont
  {Menovsky}, \citenamefont {Loidl},\ and\ \citenamefont {Saxena}}]{Dulic2001}%
  \BibitemOpen
  \bibfield  {author} {\bibinfo {author} {\bibfnamefont {D.}~\bibnamefont
  {Duli\ifmmode~\acute{c}\else \'{c}\fi{}}}, \bibinfo {author} {\bibfnamefont
  {A.}~\bibnamefont {Pimenov}}, \bibinfo {author} {\bibfnamefont
  {D.}~\bibnamefont {van~der Marel}}, \bibinfo {author} {\bibfnamefont {D.~M.}\
  \bibnamefont {Broun}}, \bibinfo {author} {\bibfnamefont {S.}~\bibnamefont
  {Kamal}}, \bibinfo {author} {\bibfnamefont {W.~N.}\ \bibnamefont {Hardy}},
  \bibinfo {author} {\bibfnamefont {A.~A.}\ \bibnamefont {Tsvetkov}}, \bibinfo
  {author} {\bibfnamefont {I.~M.}\ \bibnamefont {Sutjaha}}, \bibinfo {author}
  {\bibfnamefont {R.}~\bibnamefont {Liang}}, \bibinfo {author} {\bibfnamefont
  {A.~A.}\ \bibnamefont {Menovsky}}, \bibinfo {author} {\bibfnamefont
  {A.}~\bibnamefont {Loidl}},\ and\ \bibinfo {author} {\bibfnamefont {S.~S.}\
  \bibnamefont {Saxena}},\ }\bibfield  {title} {\bibinfo {title} {Observation
  of the transverse optical plasmon in
  ${{\mathrm{SmLa}}_{0.8}{\mathrm{Sr}}_{0.2}\mathrm{CuO}}_{4\ensuremath{-}\mathit{\ensuremath{\delta}}}$},\
  }\href {https://doi.org/10.1103/PhysRevLett.86.4144} {\bibfield  {journal}
  {\bibinfo  {journal} {Phys. Rev. Lett.}\ }\textbf {\bibinfo {volume} {86}},\
  \bibinfo {pages} {4144} (\bibinfo {year} {2001})}\BibitemShut {NoStop}%
\bibitem [{\citenamefont {Basov}\ and\ \citenamefont
  {Timusk}(2005)}]{Basov2005}%
  \BibitemOpen
  \bibfield  {author} {\bibinfo {author} {\bibfnamefont {D.~N.}\ \bibnamefont
  {Basov}}\ and\ \bibinfo {author} {\bibfnamefont {T.}~\bibnamefont {Timusk}},\
  }\bibfield  {title} {\bibinfo {title} {Electrodynamics of high-${T}_{c}$
  superconductors},\ }\href {https://doi.org/10.1103/RevModPhys.77.721}
  {\bibfield  {journal} {\bibinfo  {journal} {Rev. Mod. Phys.}\ }\textbf
  {\bibinfo {volume} {77}},\ \bibinfo {pages} {721} (\bibinfo {year}
  {2005})}\BibitemShut {NoStop}%
\bibitem [{\citenamefont {Ginzburg}\ and\ \citenamefont
  {Landau}(1950)}]{Ginzburg1950}%
  \BibitemOpen
  \bibfield  {author} {\bibinfo {author} {\bibfnamefont {V.~L.}\ \bibnamefont
  {Ginzburg}}\ and\ \bibinfo {author} {\bibfnamefont {L.~D.}\ \bibnamefont
  {Landau}},\ }\bibfield  {title} {\bibinfo {title} {On the theory of
  superconductivity},\ }\href@noop {} {\bibfield  {journal} {\bibinfo
  {journal} {Zh. Eksp. Teor. Fiz.}\ }\textbf {\bibinfo {volume} {20}},\
  \bibinfo {pages} {1064} (\bibinfo {year} {1950})}\BibitemShut {NoStop}%
\bibitem [{\citenamefont {Kogut}(1979)}]{Kogut1979}%
  \BibitemOpen
  \bibfield  {author} {\bibinfo {author} {\bibfnamefont {J.~B.}\ \bibnamefont
  {Kogut}},\ }\bibfield  {title} {\bibinfo {title} {An introduction to lattice
  gauge theory and spin systems},\ }\href
  {https://doi.org/10.1103/RevModPhys.51.659} {\bibfield  {journal} {\bibinfo
  {journal} {Rev. Mod. Phys.}\ }\textbf {\bibinfo {volume} {51}},\ \bibinfo
  {pages} {659} (\bibinfo {year} {1979})}\BibitemShut {NoStop}%
\bibitem [{\citenamefont {Yee}(1966)}]{Yee1966}%
  \BibitemOpen
  \bibfield  {author} {\bibinfo {author} {\bibfnamefont {K.}~\bibnamefont
  {Yee}},\ }\bibfield  {title} {\bibinfo {title} {Numerical solution of initial
  boundary value problems involving maxwell's equations in isotropic media},\
  }\href {https://doi.org/10.1109/TAP.1966.1138693} {\bibfield  {journal}
  {\bibinfo  {journal} {IEEE Trans. Antennas Propag.}\ }\textbf {\bibinfo
  {volume} {14}},\ \bibinfo {pages} {302} (\bibinfo {year} {1966})}\BibitemShut
  {NoStop}%
\bibitem [{\citenamefont {Denny}\ \emph {et~al.}(2015)\citenamefont {Denny},
  \citenamefont {Clark}, \citenamefont {Laplace}, \citenamefont {Cavalleri},\
  and\ \citenamefont {Jaksch}}]{Denny2015}%
  \BibitemOpen
  \bibfield  {author} {\bibinfo {author} {\bibfnamefont {S.~J.}\ \bibnamefont
  {Denny}}, \bibinfo {author} {\bibfnamefont {S.~R.}\ \bibnamefont {Clark}},
  \bibinfo {author} {\bibfnamefont {Y.}~\bibnamefont {Laplace}}, \bibinfo
  {author} {\bibfnamefont {A.}~\bibnamefont {Cavalleri}},\ and\ \bibinfo
  {author} {\bibfnamefont {D.}~\bibnamefont {Jaksch}},\ }\bibfield  {title}
  {\bibinfo {title} {Proposed parametric cooling of bilayer cuprate
  superconductors by terahertz excitation},\ }\href
  {https://doi.org/10.1103/PhysRevLett.114.137001} {\bibfield  {journal}
  {\bibinfo  {journal} {Phys. Rev. Lett.}\ }\textbf {\bibinfo {volume} {114}},\
  \bibinfo {pages} {137001} (\bibinfo {year} {2015})}\BibitemShut {NoStop}%
\bibitem [{\citenamefont {{Okamoto}}\ \emph {et~al.}(2016)\citenamefont
  {{Okamoto}}, \citenamefont {{Cavalleri}},\ and\ \citenamefont
  {{Mathey}}}]{Okamoto2016}%
  \BibitemOpen
  \bibfield  {author} {\bibinfo {author} {\bibfnamefont {J.-i.}\ \bibnamefont
  {{Okamoto}}}, \bibinfo {author} {\bibfnamefont {A.}~\bibnamefont
  {{Cavalleri}}},\ and\ \bibinfo {author} {\bibfnamefont {L.}~\bibnamefont
  {{Mathey}}},\ }\bibfield  {title} {\bibinfo {title} {Theory of enhanced
  interlayer tunneling in optically driven high-${T}_{c}$ superconductors},\
  }\href {https://doi.org/10.1103/PhysRevLett.117.227001} {\bibfield  {journal}
  {\bibinfo  {journal} {Phys. Rev. Lett.}\ }\textbf {\bibinfo {volume} {117}},\
  \bibinfo {pages} {227001} (\bibinfo {year} {2016})}\BibitemShut {NoStop}%
\bibitem [{\citenamefont {Okamoto}\ \emph {et~al.}(2017)\citenamefont
  {Okamoto}, \citenamefont {Hu}, \citenamefont {Cavalleri},\ and\ \citenamefont
  {Mathey}}]{Okamoto2017}%
  \BibitemOpen
  \bibfield  {author} {\bibinfo {author} {\bibfnamefont {J.-i.}\ \bibnamefont
  {Okamoto}}, \bibinfo {author} {\bibfnamefont {W.}~\bibnamefont {Hu}},
  \bibinfo {author} {\bibfnamefont {A.}~\bibnamefont {Cavalleri}},\ and\
  \bibinfo {author} {\bibfnamefont {L.}~\bibnamefont {Mathey}},\ }\bibfield
  {title} {\bibinfo {title} {Transiently enhanced interlayer tunneling in
  optically driven high-${T}_{c}$ superconductors},\ }\href
  {https://doi.org/10.1103/PhysRevB.96.144505} {\bibfield  {journal} {\bibinfo
  {journal} {Phys. Rev. B}\ }\textbf {\bibinfo {volume} {96}},\ \bibinfo
  {pages} {144505} (\bibinfo {year} {2017})}\BibitemShut {NoStop}%
\bibitem [{Sup()}]{Supp}%
  \BibitemOpen
  \href@noop {} {}\bibinfo {note} {{See Supplemental Material for a
  multiple-scale analysis of the sum resonance, rigidity of the Higgs time
  crystal, conductivity measurements in the time crystalline phase, and
  information on the thermal phase transition and the temperature dependence of
  the resonance frequencies}}\BibitemShut {NoStop}%
\bibitem [{\citenamefont {Landau}\ and\ \citenamefont
  {Lifšic}(1976)}]{Landau1976}%
  \BibitemOpen
  \bibfield  {author} {\bibinfo {author} {\bibfnamefont {L.~D.}\ \bibnamefont
  {Landau}}\ and\ \bibinfo {author} {\bibfnamefont {E.~M.}\ \bibnamefont
  {Lifšic}},\ }\href@noop {} {\emph {\bibinfo {title} {Mechanics}}},\ \bibinfo
  {edition} {3rd}\ ed.\ (\bibinfo  {publisher} {Butterworth-Heinemann},\
  \bibinfo {address} {Oxford},\ \bibinfo {year} {1976})\BibitemShut {NoStop}%
\bibitem [{\citenamefont {Nayfeh}(1983)}]{Nayfeh1982}%
  \BibitemOpen
  \bibfield  {author} {\bibinfo {author} {\bibfnamefont {A.~H.}\ \bibnamefont
  {Nayfeh}},\ }\bibfield  {title} {\bibinfo {title} {Combination resonances in
  the non-linear response of bowed structures to a harmonic excitation},\
  }\href {https://doi.org/10.1016/0022-460X(83)90804-0} {\bibfield  {journal}
  {\bibinfo  {journal} {J. Sound Vib.}\ }\textbf {\bibinfo {volume} {90}},\
  \bibinfo {pages} {457} (\bibinfo {year} {1983})}\BibitemShut {NoStop}%
\bibitem [{\citenamefont {{von Hoegen}}\ \emph {et~al.}(2018)\citenamefont
  {{von Hoegen}}, \citenamefont {{Mankowsky}}, \citenamefont {{Fechner}},
  \citenamefont {{F{\"o}rst}},\ and\ \citenamefont
  {{Cavalleri}}}]{vonHoegen2018}%
  \BibitemOpen
  \bibfield  {author} {\bibinfo {author} {\bibfnamefont {A.}~\bibnamefont {{von
  Hoegen}}}, \bibinfo {author} {\bibfnamefont {R.}~\bibnamefont {{Mankowsky}}},
  \bibinfo {author} {\bibfnamefont {M.}~\bibnamefont {{Fechner}}}, \bibinfo
  {author} {\bibfnamefont {M.}~\bibnamefont {{F{\"o}rst}}},\ and\ \bibinfo
  {author} {\bibfnamefont {A.}~\bibnamefont {{Cavalleri}}},\ }\bibfield
  {title} {\bibinfo {title} {Probing the interatomic potential of solids with
  strong-field nonlinear phononics},\ }\href
  {https://doi.org/10.1038/nature25484} {\bibfield  {journal} {\bibinfo
  {journal} {Nature}\ }\textbf {\bibinfo {volume} {555}},\ \bibinfo {pages}
  {79} (\bibinfo {year} {2018})}\BibitemShut {NoStop}%
\bibitem [{\citenamefont {{von Hoegen}}\ \emph {et~al.}(2019)\citenamefont
  {{von Hoegen}}, \citenamefont {{Fechner}}, \citenamefont {{F{\"o}rst}},
  \citenamefont {{Porras}}, \citenamefont {{Keimer}}, \citenamefont
  {{Michael}}, \citenamefont {{Demler}},\ and\ \citenamefont
  {{Cavalleri}}}]{vonHoegen2019}%
  \BibitemOpen
  \bibfield  {author} {\bibinfo {author} {\bibfnamefont {A.}~\bibnamefont {{von
  Hoegen}}}, \bibinfo {author} {\bibfnamefont {M.}~\bibnamefont {{Fechner}}},
  \bibinfo {author} {\bibfnamefont {M.}~\bibnamefont {{F{\"o}rst}}}, \bibinfo
  {author} {\bibfnamefont {J.}~\bibnamefont {{Porras}}}, \bibinfo {author}
  {\bibfnamefont {B.}~\bibnamefont {{Keimer}}}, \bibinfo {author}
  {\bibfnamefont {M.}~\bibnamefont {{Michael}}}, \bibinfo {author}
  {\bibfnamefont {E.}~\bibnamefont {{Demler}}},\ and\ \bibinfo {author}
  {\bibfnamefont {A.}~\bibnamefont {{Cavalleri}}},\ }\bibfield  {title}
  {\bibinfo {title} {Probing coherent charge fluctuations in
  {YBa$_2$Cu$_3$O$_{6+x}$} at wavevectors outside the light cone},\ }\href@noop
  {} {\bibfield  {journal} {\bibinfo  {journal} {arXiv e-prints}\ } (\bibinfo
  {year} {2019})},\ \Eprint {https://arxiv.org/abs/1911.08284}
  {arXiv:1911.08284 [cond-mat.supr-con]} \BibitemShut {NoStop}%
\bibitem [{\citenamefont {{Welp}}\ \emph {et~al.}(2013)\citenamefont {{Welp}},
  \citenamefont {{Kadowaki}},\ and\ \citenamefont {{Kleiner}}}]{Welp2013}%
  \BibitemOpen
  \bibfield  {author} {\bibinfo {author} {\bibfnamefont {U.}~\bibnamefont
  {{Welp}}}, \bibinfo {author} {\bibfnamefont {K.}~\bibnamefont {{Kadowaki}}},\
  and\ \bibinfo {author} {\bibfnamefont {R.}~\bibnamefont {{Kleiner}}},\
  }\bibfield  {title} {\bibinfo {title} {Superconducting emitters of {THz}
  radiation},\ }\href {https://doi.org/10.1038/nphoton.2013.216} {\bibfield
  {journal} {\bibinfo  {journal} {Nat. Photon.}\ }\textbf {\bibinfo {volume}
  {7}},\ \bibinfo {pages} {702} (\bibinfo {year} {2013})}\BibitemShut {NoStop}%
\end{thebibliography}%


\providecommand{\noopsort}[1]{}\providecommand{\singleletter}[1]{#1}%
\begin{thebibliography}{5}%
\makeatletter
\providecommand \@ifxundefined [1]{%
 \@ifx{#1\undefined}
}%
\providecommand \@ifnum [1]{%
 \ifnum #1\expandafter \@firstoftwo
 \else \expandafter \@secondoftwo
 \fi
}%
\providecommand \@ifx [1]{%
 \ifx #1\expandafter \@firstoftwo
 \else \expandafter \@secondoftwo
 \fi
}%
\providecommand \natexlab [1]{#1}%
\providecommand \enquote  [1]{``#1''}%
\providecommand \bibnamefont  [1]{#1}%
\providecommand \bibfnamefont [1]{#1}%
\providecommand \citenamefont [1]{#1}%
\providecommand \href@noop [0]{\@secondoftwo}%
\providecommand \href [0]{\begingroup \@sanitize@url \@href}%
\providecommand \@href[1]{\@@startlink{#1}\@@href}%
\providecommand \@@href[1]{\endgroup#1\@@endlink}%
\providecommand \@sanitize@url [0]{\catcode `\\12\catcode `\$12\catcode
  `\&12\catcode `\#12\catcode `\^12\catcode `\_12\catcode `\%12\relax}%
\providecommand \@@startlink[1]{}%
\providecommand \@@endlink[0]{}%
\providecommand \url  [0]{\begingroup\@sanitize@url \@url }%
\providecommand \@url [1]{\endgroup\@href {#1}{\urlprefix }}%
\providecommand \urlprefix  [0]{URL }%
\providecommand \Eprint [0]{\href }%
\providecommand \doibase [0]{https://doi.org/}%
\providecommand \selectlanguage [0]{\@gobble}%
\providecommand \bibinfo  [0]{\@secondoftwo}%
\providecommand \bibfield  [0]{\@secondoftwo}%
\providecommand \translation [1]{[#1]}%
\providecommand \BibitemOpen [0]{}%
\providecommand \bibitemStop [0]{}%
\providecommand \bibitemNoStop [0]{.\EOS\space}%
\providecommand \EOS [0]{\spacefactor3000\relax}%
\providecommand \BibitemShut  [1]{\csname bibitem#1\endcsname}%
\let\auto@bib@innerbib\@empty
\bibitem [{\citenamefont {Nayfeh}(1983)}]{Nayfeh1982}%
  \BibitemOpen
  \bibfield  {author} {\bibinfo {author} {\bibfnamefont {A.~H.}\ \bibnamefont
  {Nayfeh}},\ }\bibfield  {title} {\bibinfo {title} {Combination resonances in
  the non-linear response of bowed structures to a harmonic excitation},\
  }\href {https://doi.org/10.1016/0022-460X(83)90804-0} {\bibfield  {journal}
  {\bibinfo  {journal} {J. Sound Vib.}\ }\textbf {\bibinfo {volume} {90}},\
  \bibinfo {pages} {457} (\bibinfo {year} {1983})}\BibitemShut {NoStop}%
\bibitem [{\citenamefont {Yao}\ \emph {et~al.}(2020)\citenamefont {Yao},
  \citenamefont {Nayak}, \citenamefont {Balents},\ and\ \citenamefont
  {Zaletel}}]{Yao2020}%
  \BibitemOpen
  \bibfield  {author} {\bibinfo {author} {\bibfnamefont {N.~Y.}\ \bibnamefont
  {Yao}}, \bibinfo {author} {\bibfnamefont {C.}~\bibnamefont {Nayak}}, \bibinfo
  {author} {\bibfnamefont {L.}~\bibnamefont {Balents}},\ and\ \bibinfo {author}
  {\bibfnamefont {M.~P.}\ \bibnamefont {Zaletel}},\ }\bibfield  {title}
  {\bibinfo {title} {Classical discrete time crystals},\ }\href
  {https://doi.org/10.1038/s41567-019-0782-3} {\bibfield  {journal} {\bibinfo
  {journal} {Nat. Phys.}\ }\textbf {\bibinfo {volume} {16}},\ \bibinfo {pages}
  {438} (\bibinfo {year} {2020})}\BibitemShut {NoStop}%
\bibitem [{\citenamefont {{Hu}}\ \emph {et~al.}(2014)\citenamefont {{Hu}},
  \citenamefont {{Kaiser}}, \citenamefont {{Nicoletti}}, \citenamefont
  {{Hunt}}, \citenamefont {{Gierz}}, \citenamefont {{Hoffmann}}, \citenamefont
  {{Le Tacon}}, \citenamefont {{Loew}}, \citenamefont {{Keimer}},\ and\
  \citenamefont {{Cavalleri}}}]{Hu2014}%
  \BibitemOpen
  \bibfield  {author} {\bibinfo {author} {\bibfnamefont {W.}~\bibnamefont
  {{Hu}}}, \bibinfo {author} {\bibfnamefont {S.}~\bibnamefont {{Kaiser}}},
  \bibinfo {author} {\bibfnamefont {D.}~\bibnamefont {{Nicoletti}}}, \bibinfo
  {author} {\bibfnamefont {C.~R.}\ \bibnamefont {{Hunt}}}, \bibinfo {author}
  {\bibfnamefont {I.}~\bibnamefont {{Gierz}}}, \bibinfo {author} {\bibfnamefont
  {M.~C.}\ \bibnamefont {{Hoffmann}}}, \bibinfo {author} {\bibfnamefont
  {M.}~\bibnamefont {{Le Tacon}}}, \bibinfo {author} {\bibfnamefont
  {T.}~\bibnamefont {{Loew}}}, \bibinfo {author} {\bibfnamefont
  {B.}~\bibnamefont {{Keimer}}},\ and\ \bibinfo {author} {\bibfnamefont
  {A.}~\bibnamefont {{Cavalleri}}},\ }\bibfield  {title} {\bibinfo {title}
  {Optically enhanced coherent transport in {YBa$_{2}$Cu$_{3}$O$_{6.5}$} by
  ultrafast redistribution of interlayer coupling},\ }\href
  {https://doi.org/10.1038/nmat3963} {\bibfield  {journal} {\bibinfo  {journal}
  {Nat. Mater.}\ }\textbf {\bibinfo {volume} {13}},\ \bibinfo {pages} {705}
  (\bibinfo {year} {2014})}\BibitemShut {NoStop}%
\bibitem [{\citenamefont {Cremin}\ \emph {et~al.}(2019)\citenamefont {Cremin},
  \citenamefont {Zhang}, \citenamefont {Homes}, \citenamefont {Gu},
  \citenamefont {Sun}, \citenamefont {Fogler}, \citenamefont {Millis},
  \citenamefont {Basov},\ and\ \citenamefont {Averitt}}]{Cremin2019}%
  \BibitemOpen
  \bibfield  {author} {\bibinfo {author} {\bibfnamefont {K.~A.}\ \bibnamefont
  {Cremin}}, \bibinfo {author} {\bibfnamefont {J.}~\bibnamefont {Zhang}},
  \bibinfo {author} {\bibfnamefont {C.~C.}\ \bibnamefont {Homes}}, \bibinfo
  {author} {\bibfnamefont {G.~D.}\ \bibnamefont {Gu}}, \bibinfo {author}
  {\bibfnamefont {Z.}~\bibnamefont {Sun}}, \bibinfo {author} {\bibfnamefont
  {M.~M.}\ \bibnamefont {Fogler}}, \bibinfo {author} {\bibfnamefont {A.~J.}\
  \bibnamefont {Millis}}, \bibinfo {author} {\bibfnamefont {D.~N.}\
  \bibnamefont {Basov}},\ and\ \bibinfo {author} {\bibfnamefont {R.~D.}\
  \bibnamefont {Averitt}},\ }\bibfield  {title} {\bibinfo {title}
  {Photoenhanced metastable c-axis electrodynamics in stripe-ordered cuprate
  {La$_{1.885}$Ba$_{0.115}$CuO$_{4}$}},\ }\href
  {https://doi.org/10.1073/pnas.1908368116} {\bibfield  {journal} {\bibinfo
  {journal} {Proc. Natl. Acad. Sci. USA}\ }\textbf {\bibinfo {volume} {116}},\
  \bibinfo {pages} {19875} (\bibinfo {year} {2019})}\BibitemShut {NoStop}%
\bibitem [{\citenamefont {{von Hoegen}}\ \emph {et~al.}(2019)\citenamefont
  {{von Hoegen}}, \citenamefont {{Fechner}}, \citenamefont {{F{\"o}rst}},
  \citenamefont {{Porras}}, \citenamefont {{Keimer}}, \citenamefont
  {{Michael}}, \citenamefont {{Demler}},\ and\ \citenamefont
  {{Cavalleri}}}]{vonHoegen2019}%
  \BibitemOpen
  \bibfield  {author} {\bibinfo {author} {\bibfnamefont {A.}~\bibnamefont {{von
  Hoegen}}}, \bibinfo {author} {\bibfnamefont {M.}~\bibnamefont {{Fechner}}},
  \bibinfo {author} {\bibfnamefont {M.}~\bibnamefont {{F{\"o}rst}}}, \bibinfo
  {author} {\bibfnamefont {J.}~\bibnamefont {{Porras}}}, \bibinfo {author}
  {\bibfnamefont {B.}~\bibnamefont {{Keimer}}}, \bibinfo {author}
  {\bibfnamefont {M.}~\bibnamefont {{Michael}}}, \bibinfo {author}
  {\bibfnamefont {E.}~\bibnamefont {{Demler}}},\ and\ \bibinfo {author}
  {\bibfnamefont {A.}~\bibnamefont {{Cavalleri}}},\ }\bibfield  {title}
  {\bibinfo {title} {Probing coherent charge fluctuations in
  {YBa$_2$Cu$_3$O$_{6+x}$} at wavevectors outside the light cone},\ }\href@noop
  {} {\bibfield  {journal} {\bibinfo  {journal} {arXiv e-prints}\ } (\bibinfo
  {year} {2019})},\ \Eprint {https://arxiv.org/abs/1911.08284}
  {arXiv:1911.08284 [cond-mat.supr-con]} \BibitemShut {NoStop}%
\end{thebibliography}%

\end{document}


\title{Supplemental Material for\\Higgs time crystal in a high-$T_c$ superconductor}

\author{Guido Homann}
\affiliation{Zentrum f\"ur Optische Quantentechnologien and Institut f\"ur Laserphysik, 
	Universit\"at Hamburg, 22761 Hamburg, Germany}

\author{Jayson G. Cosme}
\affiliation{Zentrum f\"ur Optische Quantentechnologien and Institut f\"ur Laserphysik, 
	Universit\"at Hamburg, 22761 Hamburg, Germany}
\affiliation{The Hamburg Centre for Ultrafast Imaging, Luruper Chaussee 149, 22761 Hamburg, Germany}
\affiliation{National Institute of Physics, University of the Philippines, Diliman, Quezon City 1101, Philippines}

\author{Ludwig Mathey}
\affiliation{Zentrum f\"ur Optische Quantentechnologien and Institut f\"ur Laserphysik, 
	Universit\"at Hamburg, 22761 Hamburg, Germany}
\affiliation{The Hamburg Centre for Ultrafast Imaging, Luruper Chaussee 149, 22761 Hamburg, Germany}

\maketitle
\tableofcontents

\clearpage

\section{Multiple-scale analysis of the sum resonance}
Here, we derive the sum resonance of the Higgs mode and the Josephson plasmon of a monolayer cuprate superconductor in the zero-temperature limit, where the model simplifies to a 1D chain along the $c$-axis. We consider the two-mode model discussed in the main text:
\begin{align}
\partial_{t}^2 a + \gamma \partial_{t} a + \omega_{\mathrm{J}}^2 a + 2 \omega_{\mathrm{J}}^2 ah &\approx j_{\mathrm{dr}} , \\
\partial_{t}^2 h + \gamma \partial_{t} h + \omega_{\mathrm{H}}^2 h + \alpha \omega_{\mathrm{J}}^2 a^2 &\approx 0 .
\end{align}
The Higgs field is given by $h = (\psi-\psi_0)/\psi_0$ with $\psi_0$ being the equilibrium condensate amplitude, and $j_{\mathrm{dr}}$ is the current due to the drive. Note that the unitless vector potential $a$ equals the phase difference between adjacent planes in this setting.
The Higgs and plasma frequencies are $\omega_{\mathrm{H}}=\sqrt{2\mu/K \hbar^2}$ and $\omega_{\mathrm{J}}=\sqrt{t_{\mathrm{J}}/\alpha K \hbar^2}$, respectively.
Next, we expand $j_{\mathrm{dr}}$, $a$, and $h$ according to
\begin{equation}
f= f^{(0)} + \lambda f^{(1)} + \lambda^2 f^{(2)} + \mathcal{O}(\lambda^3) ,
\end{equation}
where $\lambda \ll 1$ is a small expansion parameter. Moreover, we take the driving as
\begin{equation}
j_{\mathrm{dr}}^{(1)}= j_1 \mathrm{e}^{- \mathrm{i} \omega_{\mathrm{dr}} t} + \mathrm{c.c.} ,
\end{equation}
where
\begin{equation}
\lambda |j_1|= \frac{e d \omega_{\mathrm{dr}} E_0}{\hbar \epsilon}
\end{equation}
for $E_{\mathrm{dr}} (t)= E_0 \cos(\omega_{\mathrm{dr}} t)$. From now on, we assume weak damping, that is, $\gamma= \lambda \tilde{\gamma}$. The expansion parameter $\lambda$ is also used to define multiple time scales:
\begin{equation}
T_0 \equiv t \, ,~~T_1 \equiv \lambda t .
\end{equation}
The time derivatives transform as
\begin{equation}
\partial_{t}= D_0 + \lambda D_1 + \mathcal{O}(\lambda^2) \, ,~~\partial_{t}^2= D_0^2 + 2 \lambda D_0 D_1 + \mathcal{O}(\lambda^2) ,
\end{equation}
where $D_n \equiv \frac{\partial}{\partial T_n}$. Since all the zeroth order contributions vanish, the first non-trivial contribution comes from the first order
\begin{align}
D_0^2 a^{(1)} + \omega_{\mathrm{J}}^2 a^{(1)} &= j_1 \mathrm{e}^{- \mathrm{i} \omega_{\mathrm{dr}} t} + \mathrm{c.c.} , \\
D_0^2 h^{(1)} + \omega_{\mathrm{H}}^2 h^{(1)} &= 0 .
\end{align}
This implies solutions of the form
\begin{align}
a^{(1)} &= C_{\mathrm{J}} \mathrm{e}^{- \mathrm{i} \omega_{\mathrm{J}} T_0} +  F \mathrm{e}^{- \mathrm{i} \omega_{\mathrm{dr}} T_0} + \mathrm{c.c.} , \\
h^{(1)} &= C_{\mathrm{H}} \mathrm{e}^{- \mathrm{i} \omega_{\mathrm{H}} T_0} + \mathrm{c.c.},
\end{align}
where $F$ is given by
\begin{equation}
F= \frac{j_1}{\omega_{\mathrm{J}}^2 - \omega_{\mathrm{dr}}^2}.
\end{equation}
Introducing the amplitudes $C_{\mathrm{J}}(T_1)$ and $C_{\mathrm{H}}(T_1)$ allows to describe a possible sum resonance. These amplitudes are determined in the following. In second order, we have
\begin{align}
D_0^2 a^{(2)} + \omega_{\mathrm{J}}^2 a^{(2)} &= -2 D_0 D_1 a^{(1)} - \tilde{\gamma} D_0 a^{(1)} - 2 \omega_{\mathrm{J}}^2 a^{(1)} h^{(1)} , \\
D_0^2 h^{(2)} + \omega_{\mathrm{H}}^2 h^{(2)} &= -2 D_0 D_1 h^{(1)} - \tilde{\gamma} D_0 h^{(1)} - \alpha \omega_{\mathrm{J}}^2 [a^{(1)}]^2 .
\end{align}
Substituting the first order solutions into the second order equations leads to
\begin{align}
\begin{split}
D_0^2 a^{(2)} + \omega_{\mathrm{J}}^2 a^{(2)} ={}& \mathrm{i} (2 D_1 + \tilde{\gamma}) \bigl( \omega_{\mathrm{J}} C_{\mathrm{J}} \mathrm{e}^{- \mathrm{i} \omega_{\mathrm{J}} T_0} + \omega_{\mathrm{dr}} F \mathrm{e}^{- \mathrm{i} \omega_{\mathrm{dr}} T_0} \bigr) \\
& - 2 \omega_{\mathrm{J}}^2 \Bigl( C_{\mathrm{J}} C_{\mathrm{H}} \mathrm{e}^{- \mathrm{i}(\omega_{\mathrm{J}} + \omega_{\mathrm{H}}) T_0} + C_{\mathrm{J}} C_{\mathrm{H}}^* \mathrm{e}^{- \mathrm{i}(\omega_{\mathrm{J}} - \omega_{\mathrm{H}}) T_0} + F C_{\mathrm{H}} \mathrm{e}^{- \mathrm{i}(\omega_{\mathrm{dr}} + \omega_{\mathrm{H}}) T_0} + F C_{\mathrm{H}}^* \mathrm{e}^{- \mathrm{i}(\omega_{\mathrm{dr}} - \omega_{\mathrm{H}}) T_0} \Bigr) + \mathrm{c.c.} ,
\end{split} \\
\begin{split}
D_0^2 h^{(2)} + \omega_{\mathrm{H}}^2 h^{(2)} ={}& \mathrm{i} (2 D_1 + \tilde{\gamma}) \omega_{\mathrm{H}} C_{\mathrm{H}} \mathrm{e}^{- \mathrm{i} \omega_{\mathrm{H}} T_0} - \alpha \omega_{\mathrm{J}}^2 \Bigl( |C_{\mathrm{J}}|^2 + |F|^2 + C_{\mathrm{J}}^2 \mathrm{e}^{- 2 \mathrm{i} \omega_{\mathrm{J}} T_0} + F^2 \mathrm{e}^{- 2 \mathrm{i} \omega_{\mathrm{dr}} T_0} \\
& \qquad \qquad \qquad \qquad \qquad \qquad \qquad + 2 F C_{\mathrm{J}} \mathrm{e}^{- \mathrm{i}(\omega_{\mathrm{dr}} + \omega_{\mathrm{J}}) T_0} + 2 F C_{\mathrm{J}}^* \mathrm{e}^{- \mathrm{i}(\omega_{\mathrm{dr}} - \omega_{\mathrm{J}}) T_0} \Bigr) + \mathrm{c.c.} .
\end{split}
\end{align}
To study the behavior near the sum resonance, we write
\begin{equation}\label{eq:sum}
\omega_{\mathrm{dr}}= \omega_{\mathrm{J}} + \omega_{\mathrm{H}} + \lambda \delta
\end{equation}
with the detuning $\delta$. Inserting this into the second order equations induces secular terms, which we demand to vanish:
\begin{align}
\mathrm{i} (2 D_1 + \tilde{\gamma}) \omega_{\mathrm{J}} C_{\mathrm{J}} - 2 \omega_{\mathrm{J}}^2 F C_{\mathrm{H}}^* \mathrm{e}^{- \mathrm{i} \delta T_1} &= 0 , \label{eq:sec1} \\
\mathrm{i} (2 D_1 + \tilde{\gamma}) \omega_{\mathrm{H}} C_{\mathrm{H}} - 2 \alpha \omega_{\mathrm{J}}^2 F C_{\mathrm{J}}^* \mathrm{e}^{- \mathrm{i} \delta T_1} &= 0 . \label{eq:sec2}
\end{align}
The conditions \eqref{eq:sec1} and \eqref{eq:sec2} imply solutions of the form
\begin{align}
C_{\mathrm{J}} &= \tilde{C_{\mathrm{J}}} \mathrm{e}^{(r - \mathrm{i} \delta) T_1} , \\
C_{\mathrm{H}} &= \tilde{C_{\mathrm{J}}} \mathrm{e}^{r^* T_1}.
\end{align}
Using this ansatz, we find
\begin{equation}
r= -\frac{(\tilde{\gamma} - \mathrm{i} \delta)}{2} \pm \frac{1}{2} \sqrt{\frac{4 \alpha \omega_{\mathrm{J}}^3}{\omega_{\mathrm{H}} } |F|^2 - \delta^2} .
\end{equation}
If the real part of $r$ is positive, the amplitudes $C_{\mathrm{J}}$ and $C_{\mathrm{H}}$ grow exponentially. Such a behaviour signals the excitation of the sum resonance. It requires a sufficient driving amplitude given by the condition
\begin{equation}
|F|^2 > \bigl( \tilde{\gamma}^2 + \delta^2 \bigr) \frac{\omega_{\mathrm{H}}}{4 \alpha \omega_{\mathrm{J}}^3} .
\end{equation}
Let us consider the case $\omega_{\mathrm{dr}}= \omega_{\mathrm{J}} + \omega_{\mathrm{H}}$, i.e., $\delta=0$. In this case, the required driving amplitude to induce the sum resonance is
\begin{equation}
E_0 > \gamma \sqrt{ \frac{2 n_0 K \hbar^2}{\epsilon \epsilon_0} } \biggl( \frac{2 \omega_{\mathrm{J}} + \omega_{\mathrm{H}}}{\omega_{\mathrm{J}} + \omega_{\mathrm{H}}} \biggr) \biggl( \frac{\omega_{\mathrm{H}}}{\omega_{\mathrm{J}}} \biggr)^{3/2} \approx 8 \times 10^{-3}~\mathrm{MV \, cm^{-1}} 
\end{equation}
for the parameters specified in the main text. Higher order terms play an important role in saturating the amplitude of oscillations \cite{Nayfeh1982}, which can be understood from the perspective of non-linear oscillators having amplitude dependent eigenfrequencies.

In the case of driving close to the difference frequency,
\begin{equation}
\omega_{\mathrm{dr}}= \omega_{\mathrm{J}} - \omega_{\mathrm{H}} + \lambda \delta ,
\end{equation}
we find
\begin{equation}
r= - \frac{(\tilde{\gamma} - \mathrm{i} \delta)}{2} \pm \frac{1}{2} \sqrt{- \frac{4 \alpha \omega_{\mathrm{J}}^3}{\omega_{\mathrm{H}} } |F|^2 - \delta^2} .
\end{equation}
Here the real part of $r$ is always negative. Hence, there is no difference resonance in the system.

\section{Rigidity of the Higgs time crystal}
The following zero-temperature simulations refer to the bilayer cuprate superconductor specified in the main text. We take the driving as
\begin{equation}
E_{\mathrm{dr}}(t)= \frac{E_0}{2} \mathrm{cos}(\omega_{\mathrm{dr}} t) \biggl[1+ \mathrm{tanh} \biggl(\frac{t}{\tau} \biggr) \biggr] ,
\end{equation}
where $E_0$ is the strength of the external field effectively penetrating the sample. Additionally, the external drive is characterised by the frequency $\omega_{\mathrm{dr}}$ and the rise time $\tau$.

\begin{figure}[!h]
	\centering
	\includegraphics[scale=1]{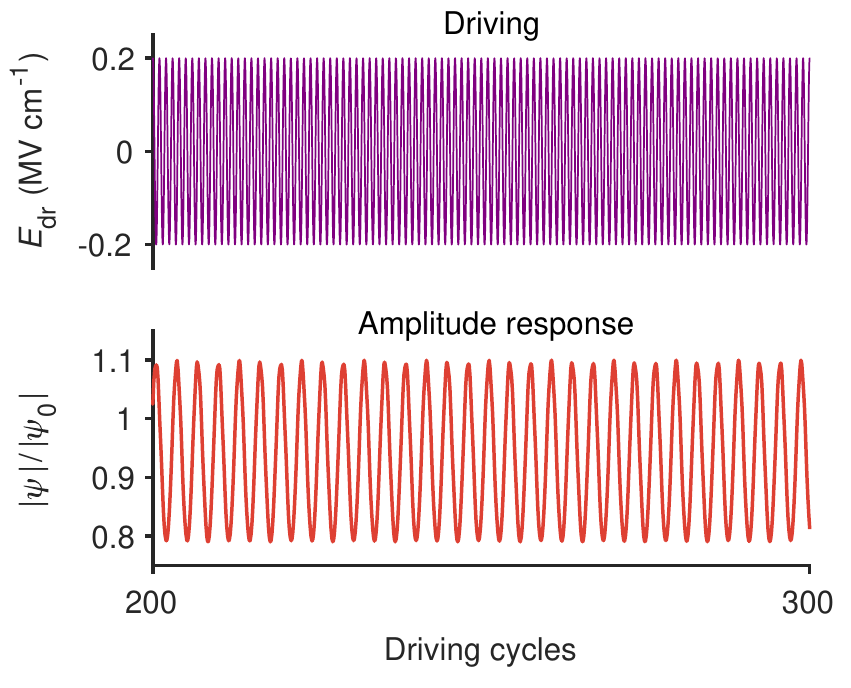}\llap{\parbox[b]{17cm}{\textcolor{black}{(a)}\\\rule{0ex}{6.8cm}}}
	\hfill
	\includegraphics[scale=1]{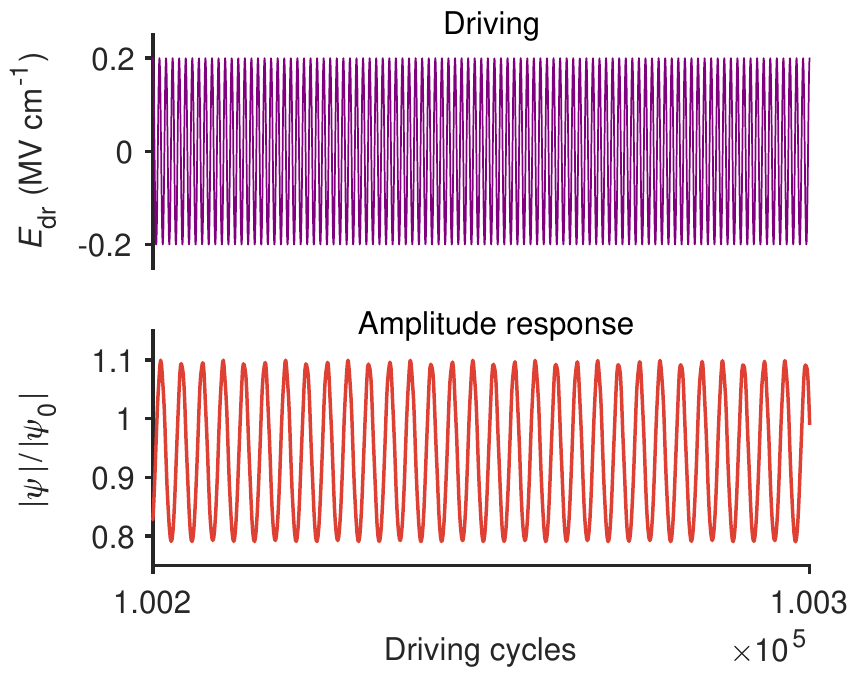}\llap{\parbox[b]{17.1cm}{\textcolor{black}{(b)}\\\rule{0ex}{6.8cm}}}
	\caption{Long-time persistence of the subharmonic response at $T=0$. (a) Amplitude response after 200 driving cycles. (b) Amplitude response after $1.002 \times 10^5$ driving cycles. }
	\label{fig:longtime} 
\end{figure}

To realise the sum resonance of the Higgs mode and the upper Josephson plasmon, we drive the electric field with $E_0= 0.2~\mathrm{MV \, cm^{-1}}$ and $\omega_{\mathrm{dr}}/2\pi= 21~\mathrm{THz}$.
The long-time persistence of the time-translation symmetry breaking is exemplified in Fig.~\ref{fig:longtime}, where the subharmonic oscillations in the condensate amplitude are found to survive even after $10^5$ driving cycles.

\begin{figure}[!h]
	\centering
	\includegraphics[scale=1]{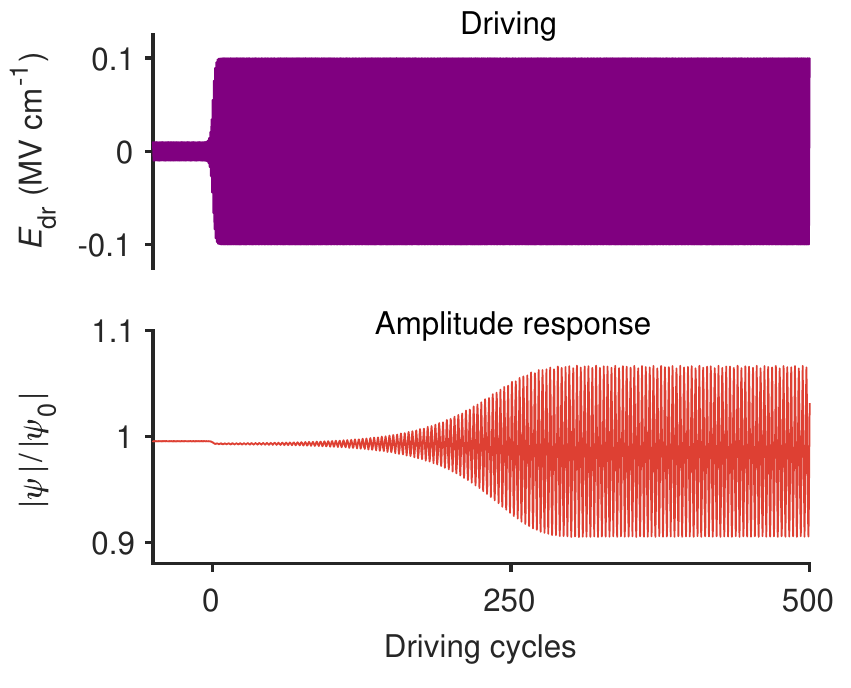}\llap{\parbox[b]{17cm}{\textcolor{black}{(a)}\\\rule{0ex}{6.8cm}}}
	\hfill
	\includegraphics[scale=1]{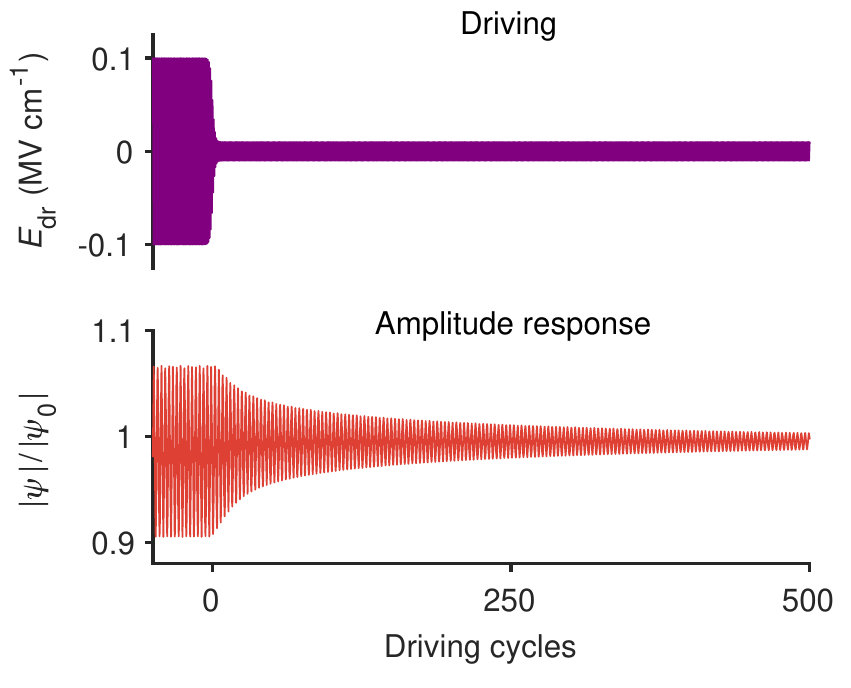}\llap{\parbox[b]{17.1cm}{\textcolor{black}{(b)}\\\rule{0ex}{6.8cm}}}
	\caption{Dynamical transitions between normal and time crystalline phase. (a) Transition from normal to time crystalline phase. (b) Transition from time crystalline to normal phase.}
	\label{fig:dynTrans} 
\end{figure}

As discussed in Ref.~\cite{Yao2020}, a signature of a phase transition to a time crystalline order in classical systems is the hysteretic behaviour across a critical point. Here, we demonstrate an indicator of such hysteresis in the response of the condensate amplitude. This can be seen in Fig.~\ref{fig:dynTrans} as we tune the driving amplitude across the time crystal-normal response transition from $E_0= 0.08~\mathrm{MV \, cm^{-1}}$  to $E_0= 0.1~\mathrm{MV \, cm^{-1}}$ and vice versa, while keeping the driving frequency fixed at $\omega_{\mathrm{dr}}/2\pi= 21~\mathrm{THz}$. In particular, there is a clear difference in the time that it takes the system to enter and leave the time crystalline phase.

\clearpage
\section{Optical conductivity of the Higgs time crystal}
\begin{figure}[!h]
	\centering
	\includegraphics[scale=1]{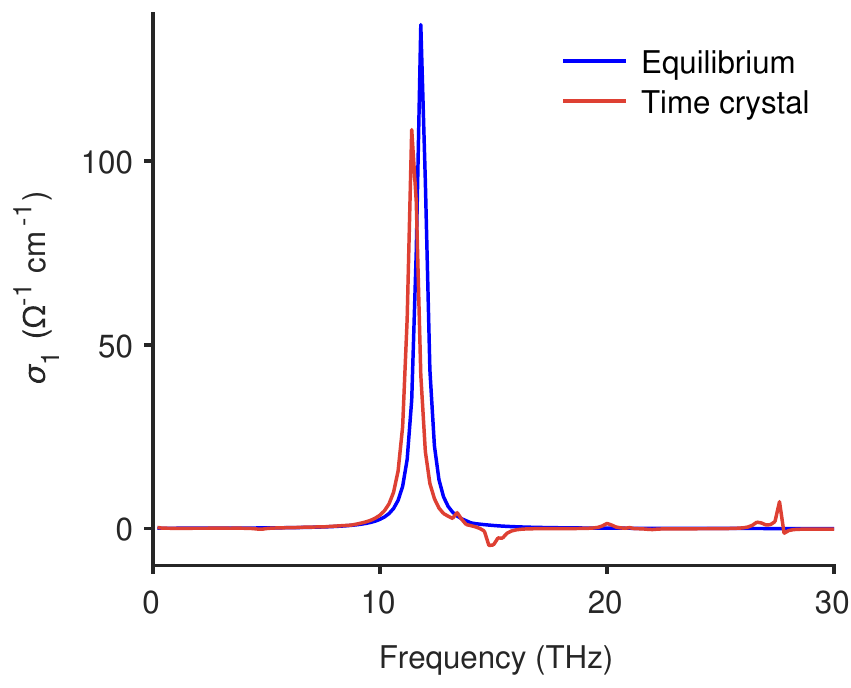}\llap{\parbox[b]{17cm}{\textcolor{black}{(a)}\\\rule{0ex}{6.6cm}}}
	\hfill
	\includegraphics[scale=1]{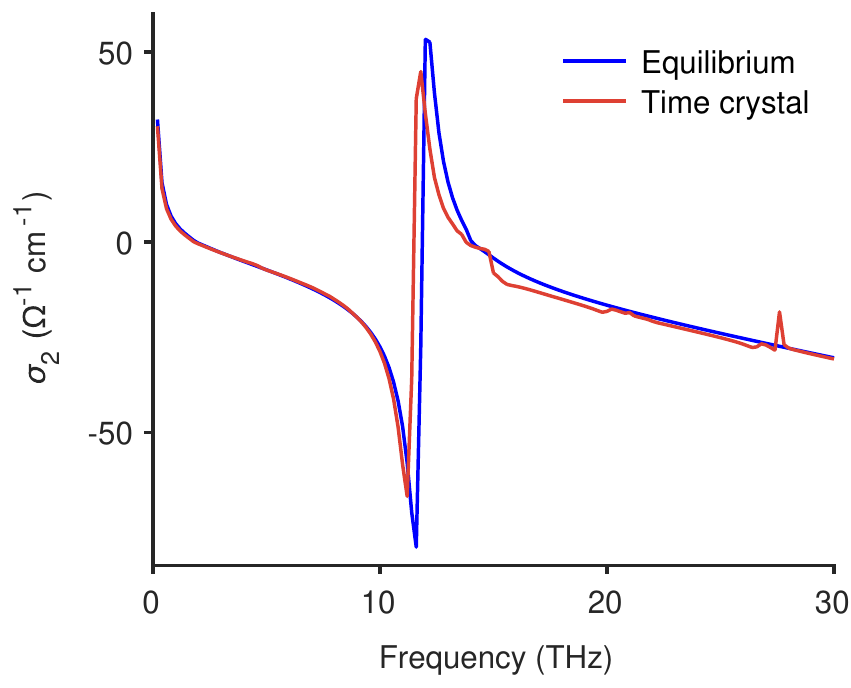}\llap{\parbox[b]{17.1cm}{\textcolor{black}{(b)}\\\rule{0ex}{6.6cm}}}
	\caption{Optical conductivity in the time crystalline phase. (a) Real part of the optical conductivity. (b) Imaginary part of the optical conductivity.}
	\label{fig:conductivity} 
\end{figure}

The optical conductivity $\sigma(\omega)$ is a crucial quantity to characterise the electric transport properties of a superconductor in the linear response regime. It is a macroscopic observable that can be measured in pump-probe experiments \cite{Hu2014, Cremin2019}. Here, we investigate how the emergence of the Higgs time crystal alters the $c$-axis optical conductivity of a bilayer cuprate superconductor with parameters as specified in the main text. For this purpose, the system is driven into the time crystalline phase with $E_0= 0.2~\mathrm{MV \, cm^{-1}}$ and $\omega_{\mathrm{dr}}/2\pi= 21~\mathrm{THz}$ at $T=0$. Then, we add a probing term to the external drive, 
\begin{equation}
E_{\mathrm{dr}}(t)= E_0 \mathrm{cos}(\omega_{\mathrm{dr}} t) + \frac{E_{\mathrm{pr}}}{2} \mathrm{cos}(\omega_{\mathrm{pr}} t) \biggl[1+ \mathrm{tanh} \biggl(\frac{\omega_{\mathrm{pr}} (t-t_{\mathrm{pr}})}{2\pi} \biggr) \biggr] ,
\end{equation}
where $t_{\mathrm{pr}}= 10$ ps. The probing amplitude $E_{\mathrm{pr}}$ has to be one order of magnitude smaller than $E_0$ to enter the linear response regime. We evaluate $\sigma(\omega)= J_{\mathrm{tot}}(\omega)/E(\omega)$ from a Fourier analysis over 50 ps in the steady state. The average electric field along the $c$-axis is given by
\begin{equation}
E(t)= \frac{d_s \overline{E_s(t)} + d_w \overline{E_w(t)}}{d_s + d_w} ,
\end{equation}
where $\overline{E_{s,w}(t)}$ denotes the spatial average of electric fields along either strong or weak junctions.
The total current is the sum of the average supercurrent and the average displacement current inside the sample, that is,
\begin{equation}
J_{\mathrm{tot}}(t)= J(t) + \frac{d_s \epsilon_s \epsilon_0 \overline{\partial_{t} E_s} + d_w  \epsilon_w \epsilon_0 \overline{\partial_{t} E_w}}{d_s + d_w},
\end{equation}
where $J(t)$ is the supercurrent given in the main text.
As visible in Fig.~\ref{fig:conductivity}(a), the real part of the optical conductivity acquires additional resonance peaks in the time crystalline phase, especially at $\omega_{\mathrm{L}}= \omega_{\mathrm{dr}} - \omega_{\mathrm{H}}$ and $\omega_{\mathrm{R}}= \omega_{\mathrm{dr}} + \omega_{\mathrm{H}}$. These frequencies correspond to the side peaks previously observed in the supercurrent spectra. Remarkably, the current response is amplified at the left side peak while attenuated at the right side peak. The counterparts of the peaks in $\sigma_1$ are sharp edges in $\sigma_2$ as evidenced by Fig.~\ref{fig:conductivity}(b). The depletion of the condensate tends to reduce the plasma frequencies in the time crystalline phase. This effect is most apparent for the transverse Josephson plasmon shifting from 11.8 THz to 11.4 THz. For the same reason, we find a smaller prefactor of the $1/\omega$ divergence of $\sigma_2$ at low frequencies.

\clearpage
\section{Thermal phase transition}
\begin{figure}[!h]
	\centering
	\includegraphics[scale=1]{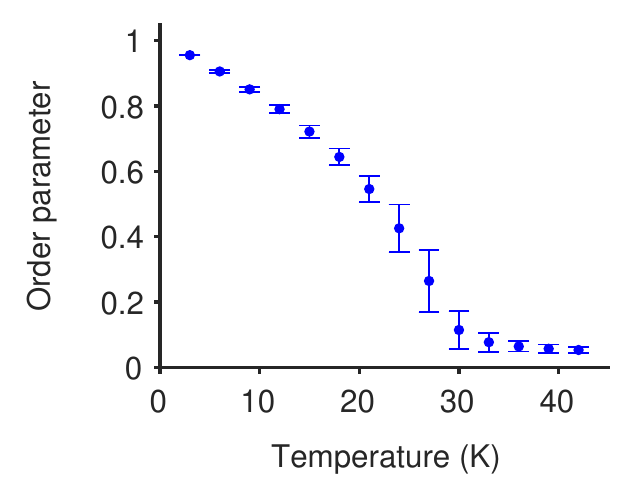}\llap{\parbox[b]{12.5cm}{\textcolor{black}{(a)}\\\rule{0ex}{4.6cm}}}
	\includegraphics[scale=1]{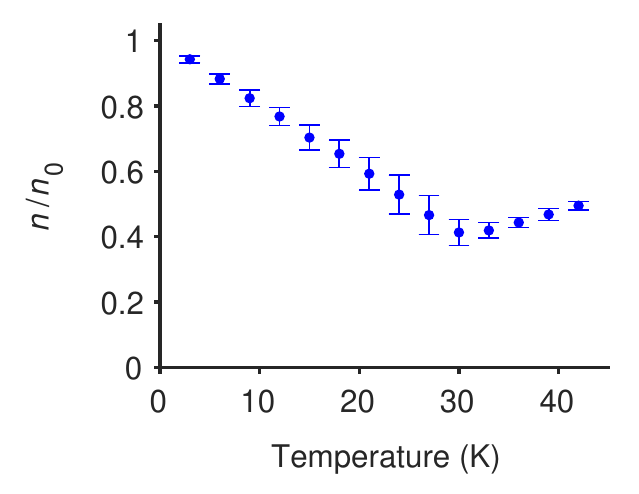}\llap{\parbox[b]{12.5cm}{\textcolor{black}{(b)}\\\rule{0ex}{4.6cm}}} \\
	\includegraphics[scale=1]{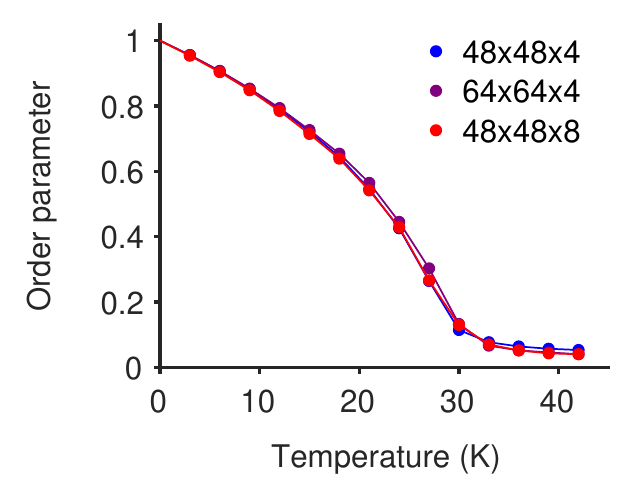}\llap{\parbox[b]{12.5cm}{\textcolor{black}{(c)}\\\rule{0ex}{4.6cm}}}
	\includegraphics[scale=1]{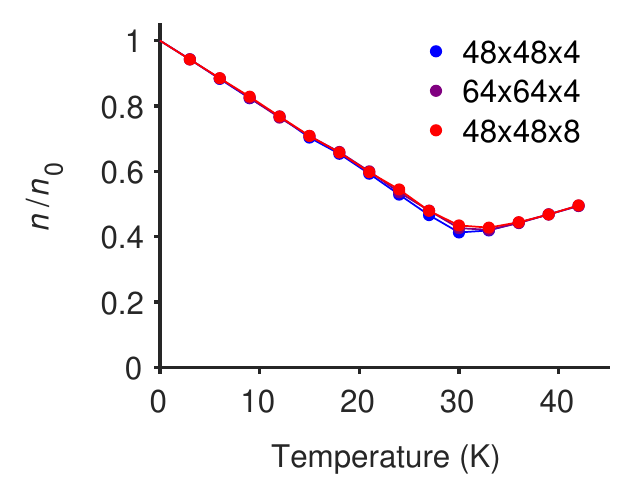}\llap{\parbox[b]{12.5cm}{\textcolor{black}{(d)}\\\rule{0ex}{4.6cm}}}
	\caption{Phase transition of a bilayer cuprate superconductor. (a) Order parameter for a system of $48\times 48 \times 4$ lattice sites. (b) Condensate density for a system of $48\times 48 \times 4$ lattice sites. The error bars indicate the standard deviations of the ensemble averages. (c) Order parameter for various system sizes. (d) Condensate density for various system sizes. The physical parameters are the same as for the bilayer system considered in the main text.}
	\label{fig:phaseTrans} 
\end{figure}

Here, we elaborate on the thermal phase transition of the simulated bilayer cuprate superconductor, see main text for parameters. The thermal equilibrium at a given temperature is established as follows. We initialise the system in its ground state at $T=0$ and let the dynamics evolve without external driving, influenced only by thermal fluctuations and dissipation. After 10 ps, the average condensate density $n= \frac{1}{N} \sum_{\mathbf{r}} |\psi_{\mathbf{r}}|^2$ and the phase coherence are converged, indicating thermal equilbrium. To characterise the phase transition, we introduce the order parameter
\begin{equation}
O= \frac{1}{n} \bigg|\frac{1}{N/2} \sum_{l,m,n \in \mathrm{odd}} \psi_{l,m,n+1}^* \psi_{l,m,n} \, \mathrm{e}^{\mathrm{i} a_{l,m,n}^z} \bigg| .
\end{equation}
The order parameter measures the phase coherence of the condensate across different bilayers. For each trajectory, it is evaluated from the average of 200 measurements within a time interval of 2 ps. Finally, we take the ensemble average of 100 trajectories. As shown in Fig.~\ref{fig:phaseTrans}(a), the temperature dependence of the order parameter is reminiscent of a second order phase transition. Due to the finite size of the simulated system, the order parameter converges to a plateau with non-zero value for high temperatures. Instead of a sharp discontinuity, one finds a distinct crossover at $T_c \sim 30$ K. We also note that the lower Josephson plasmon vanishes in this temperature regime, which agrees with experimental observations \cite{vonHoegen2019}. Figure~\ref{fig:phaseTrans}(b) reveals that the condensate density does not drop below $0.4 \, n_0$ through the phase transition. Strikingly, the condensate density decreases almost linearly with temperature below $T_c$. By contrast, it undergoes a nearly linear increase above the transition temperature. We see in Figs.~\ref{fig:phaseTrans}(c) and \ref{fig:phaseTrans}(d) that the phase transition is only weakly modified by increasing the system size.

\clearpage
\section{Temperature dependence of the resonance frequencies}
\begin{figure}[!b]
	\centering
	\includegraphics[scale=1]{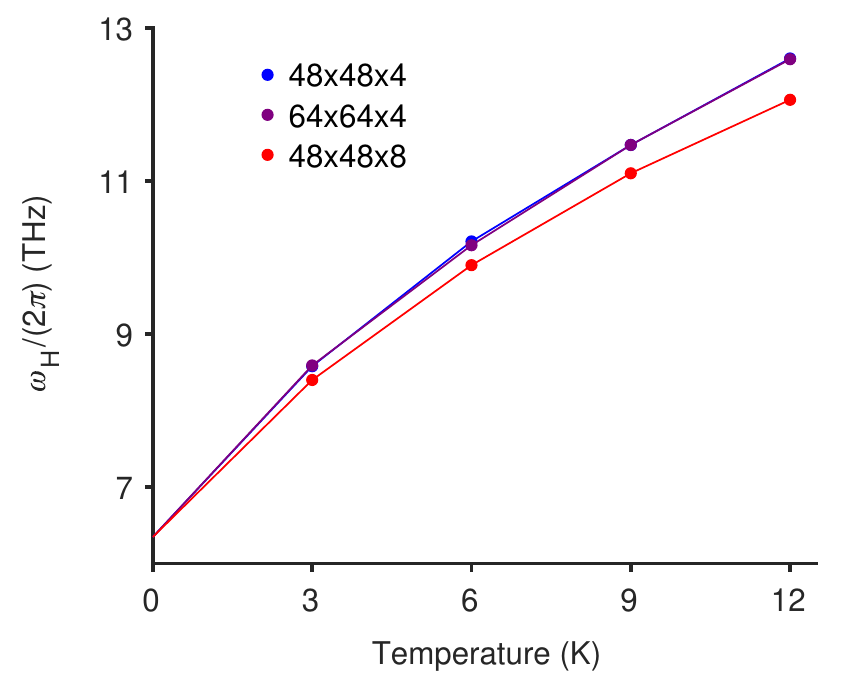}\llap{\parbox[b]{17cm}{\textcolor{black}{(a)}\\\rule{0ex}{6.6cm}}}
	\hfill
	\includegraphics[scale=1]{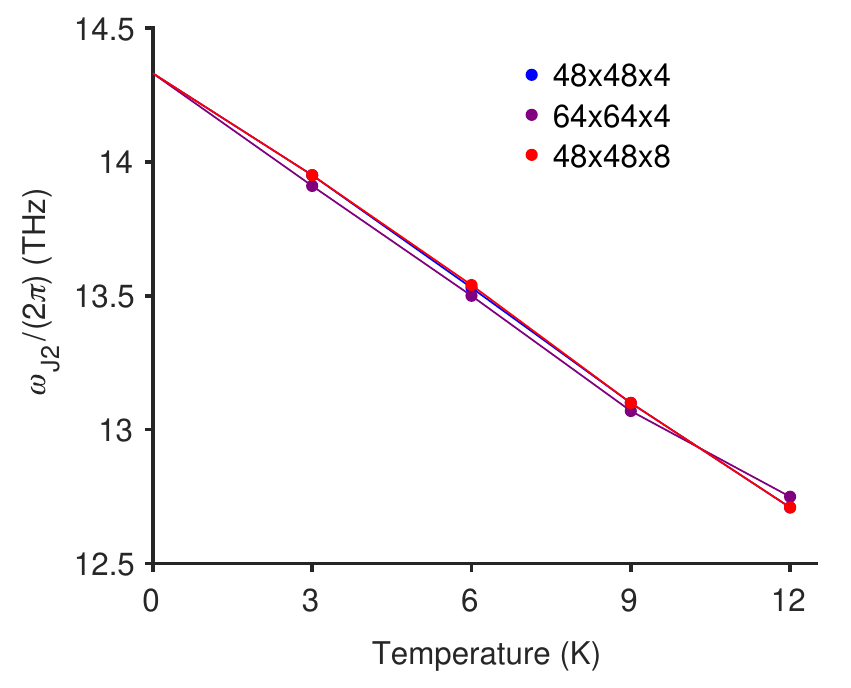}\llap{\parbox[b]{17.1cm}{\textcolor{black}{(b)}\\\rule{0ex}{6.6cm}}}
	\caption{Temperature dependence of the the resonance frequencies. (a) Temperature dependence of the Higgs frequency. (b) Temperature dependence of the upper Josephson plasma frequency.}
	\label{fig:tempResonances} 
\end{figure}

In this section, we discuss the temperature dependence of the resonance frequencies, which can be deduced from the undriven dynamics of the superconductor in thermal equilibrium. More precisely, the Higgs mode and the longitudinal Josephson plasmons appear as peaks in the amplitude and supercurrent spectra, respectively. We fit a Lorentzian to the corresponding maxima in the thermal spectra (ensemble average of 100 trajectories).
Figure~\ref{fig:tempResonances} displays the temperature dependence of the Higgs and upper Josephson plasma frequencies up to $\sim0.4 \, T_c$ for various system sizes.

While the upper Josephson plasmon is weakly shifted towards smaller frequencies, the Higgs frequency notably increases with temperature. The temperature dependence of the Higgs frequency does not significantly depend on the system size, but on the in-plane tunnelling $t_{ab}$ as inferred from additional simulations. Additionally, a comparison to thermal spectra with different $t_{ab}$ indicates a minor role of the condensate density in this process. That is why our analysis is focused on the correction of the Higgs frequency arising from fourth order coupling terms between the Higgs field and the vector potential. Due to the dominant role of the in-plane dynamics in this process, we perform the following calculations in 2D. Expanding around the thermal equilibrium at a given temperature yields
\begin{equation} \label{eq:eom_finiteT}
\partial_{t}^2 h_{\mathbf{r}} + \gamma \partial_{t} h_{\mathbf{r}} + \frac{2 \mu}{K \hbar^2} h_{\mathbf{r}} + \frac{t_{ab}}{K \hbar^2} \sum_{\mathbf{r'} \in \mathrm{NN}} (h_{\mathbf{r}} - h_{\mathbf{r'}}) + \frac{t_{ab}}{2K \hbar^2} \sum_{\mathbf{r'} \in \mathrm{NN}} \theta_{\mathbf{rr'}}^2 h_{\mathbf{r'}} \approx 0 ,
\end{equation}
where the sum is restricted to the nearest neighbours (NN) of $\mathbf{r}$ in the $ab$-plane, and $\theta_{\mathbf{rr'}}= a_{\mathbf{rr'}} + \mathrm{arg}(\psi_{\mathbf{r}}) - \mathrm{arg}(\psi_{\mathbf{r'}})$ denotes the gauge-invariant phase between neighbouring sites. The notation $a_{\mathbf{rr'}}$ means the bond-directed component of the vector potential at $(\mathbf{r} + \mathbf{r'})/2$ with $a_{\mathbf{r'r}}= -a_{\mathbf{rr'}}$.
A Fourier transform leads to
\begin{equation}
\sum_{\mathbf{k}} \Bigl(\partial_{t}^2 h_{\mathbf{k}} + \gamma \partial_{t} h_{\mathbf{k}} + \omega_{\mathrm{H}}^2(\mathbf{k}) h_{\mathbf{k}} \Bigr) \mathrm{e}^{\mathrm{i} \mathbf{k} \cdot \mathbf{r}} \approx - \frac{t_{ab}}{K \hbar^2 M} \sum_{\mathbf{p},\mathbf{q}} \biggl[ \mathrm{cos}\biggl(\frac{p_x + 2q_x}{2} d_{ab} \biggr) (\theta_x^2)_{\mathbf{p}} + \mathrm{cos}\biggl(\frac{p_y + 2q_y}{2} d_{ab} \biggr) (\theta_y^2)_{\mathbf{p}} \biggr] h_{\mathbf{q}} \mathrm{e}^{\mathrm{i} (\mathbf{p}+\mathbf{q}) \cdot \mathbf{r}} ,
\end{equation}
where $M$ is the total number of sites in the $ab$-plane, and
\begin{equation}
\omega_{\mathrm{H}}^2(\mathbf{k})= \frac{2 \mu}{K \hbar^2} + \frac{2t_{ab}}{K \hbar^2} \Bigl[2- \mathrm{cos}(k_x d_{ab}) - \mathrm{cos}(k_y d_{ab}) \Bigr] .
\end{equation}
The equation of motion for a given momentum mode reads
\begin{equation}
\partial_{t}^2 h_{\mathbf{k}} + \gamma \partial_{t} h_{\mathbf{k}} + \omega_{\mathrm{H}}^2(\mathbf{k}) h_{\mathbf{k}} \approx - \frac{t_{ab}}{K \hbar^2 M} \sum_{\mathbf{q}} \biggl[ \mathrm{cos}\biggl(\frac{k_x + q_x}{2} d_{ab} \biggr) (\theta_x^2)_{\mathbf{k}-\mathbf{q}} + \mathrm{cos}\biggl(\frac{k_y + q_y}{2} d_{ab} \biggr) (\theta_y^2)_{\mathbf{k}-\mathbf{q}} \biggr] h_{\mathbf{q}} .
\end{equation}
So, we have
\begin{equation} \label{eq:eom_Fourier}
\partial_{t}^2 h_0 + \gamma \partial_{t} h_0 + \omega_{\mathrm{H}}^2(0) h_0 \approx - \frac{t_{ab}}{K \hbar^2} \frac{(\theta_x^2)_0 + (\theta_y^2)_0}{M} h_0 - \frac{t_{ab}}{K \hbar^2 M} \sum_{\mathbf{q} \neq 0} \biggl[ \mathrm{cos}\biggl(\frac{q_x d_{ab}}{2} \biggr) (\theta_x^2)_{-\mathbf{q}} + \mathrm{cos}\biggl(\frac{q_y d_{ab}}{2} \biggr) (\theta_y^2)_{-\mathbf{q}} \biggr] h_{\mathbf{q}} .
\end{equation}

\begin{figure}[!t]
	\centering
	\includegraphics[scale=1]{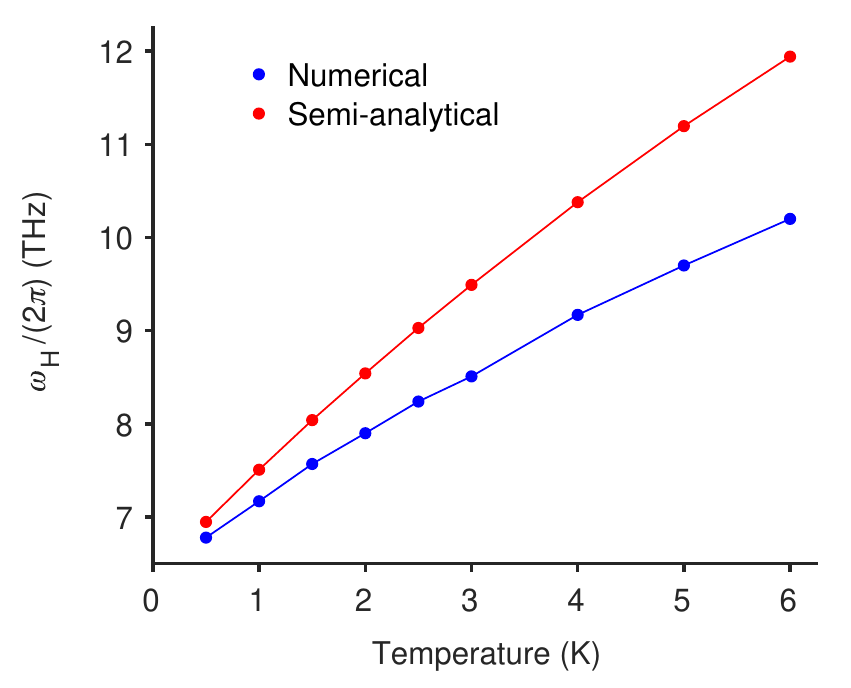}
	\caption{Increase of the Higgs frequency at low temperatures. The numerical results are compared to a semi-analytical estimate.}
	\label{fig:predHiggs} 
\end{figure}
\noindent
To determine the finite-momentum modes on the right-hand side of equation~\eqref{eq:eom_Fourier}, we apply a rotating wave approximation:
\begin{equation} \label{eq:rwa}
h_{\mathbf{q}} \approx - \frac{t_{ab}}{K \hbar^2 \omega_{\mathrm{H}}^2 (\mathbf{q}) M} \sum_{\mathbf{k}} \biggl[ \mathrm{cos}\biggl(\frac{k_x + q_x}{2} d_{ab} \biggr) (\theta_x^2)_{\mathbf{q}-\mathbf{k}} + \mathrm{cos}\biggl(\frac{k_y + q_y}{2} d_{ab} \biggr) (\theta_y^2)_{\mathbf{q}-\mathbf{k}} \biggr] h_{\mathbf{k}} .
\end{equation}
Furthermore, we assume that the zero-momentum mode provides the main contribution to the sum in equation~\eqref{eq:rwa}, leading to
\begin{equation} \label{eq:zeroModeApprox}
h_{\mathbf{q}} \approx - \frac{t_{ab}}{K \hbar^2 \omega_{\mathrm{H}}^2 (\mathbf{q}) M} \biggl[ \mathrm{cos}\biggl(\frac{q_x d_{ab}}{2} \biggr) (\theta_x^2)_{\mathbf{q}} + \mathrm{cos}\biggl(\frac{q_y  d_{ab}}{2} \biggr) (\theta_y^2)_{\mathbf{q}} \biggr] h_0 .
\end{equation}
Inserting this into equation~\eqref{eq:eom_Fourier} yields
\begin{align}
\begin{split} \label{eq:eom_approx}
& \partial_{t}^2 h_0 + \gamma \partial_{t} h_0 + \omega_{\mathrm{H}}^2(0) h_0 \\
& \approx - \frac{t_{ab}}{K \hbar^2} \frac{(\theta_x^2)_0 + (\theta_y^2)_0}{M} h_0 + \frac{t_{ab}^2}{K^2 \hbar^4} \sum_{\mathbf{q} \neq 0} \biggl[ \frac{F_x({\mathbf{q}}) (\theta_x^2)_{\mathbf{q}}(\theta_x^2)_{-\mathbf{q}}} {M^2} + \frac{F_y({\mathbf{q}}) (\theta_y^2)_{\mathbf{q}}(\theta_y^2)_{-\mathbf{q}}} {M^2} + \frac{2F_{xy}({\mathbf{q}}) (\theta_x^2)_{\mathbf{q}}(\theta_y^2)_{-\mathbf{q}}} {M^2} \biggr] h_0 ,
\end{split}
\end{align}
where
\begin{align}
F_x(\mathbf{q}) &= \frac{1}{\omega_{\mathrm{H}}^2(\mathbf{q})} \, \mathrm{cos}^2\biggl(\frac{q_x d_{ab}}{2} \biggr) , \\
F_y(\mathbf{q}) &= \frac{1}{\omega_{\mathrm{H}}^2(\mathbf{q})} \, \mathrm{cos}^2\biggl(\frac{q_y d_{ab}}{2} \biggr) , \\
F_{xy}(\mathbf{q}) &= \frac{1}{\omega_{\mathrm{H}}^2(\mathbf{q})} \, \mathrm{cos}\biggl(\frac{q_x d_{ab}}{2} \biggr) \mathrm{cos}\biggl(\frac{q_y d_{ab}}{2} \biggr) .
\end{align}
This implies the temperature-dependent Higgs frequency
\begin{equation} \label{eq:omegaH_temp}
\omega_{\mathrm{H}} (\mathbf{k}=0, T)= \omega_{\mathrm{H}}(0,0) \,  \sqrt{1 + \Delta_1(T) + \Delta_2(T)} ,
\end{equation}
with the corrections
\begin{align}
\Delta_1(T) &= \frac{t_{ab}}{2 \mu} \biggl\langle \frac{(\theta_x^2)_0 + (\theta_y^2)_0}{M} \biggr\rangle , \\
\Delta_2(T) &= - \frac{t_{ab}^2}{2 \mu K \hbar^2} \sum_{\mathbf{q} \neq 0} \biggl\langle \frac{F_x({\mathbf{q}}) (\theta_x^2)_{\mathbf{q}}(\theta_x^2)_{-\mathbf{q}}} {M^2} + \frac{F_y({\mathbf{q}}) (\theta_y^2)_{\mathbf{q}}(\theta_y^2)_{-\mathbf{q}}} {M^2} + \frac{2F_{xy}({\mathbf{q}}) (\theta_x^2)_{\mathbf{q}}(\theta_y^2)_{-\mathbf{q}}} {M^2} \biggr\rangle .
\end{align}
The estimate in Eq.~\eqref{eq:omegaH_temp} is compared to the purely numerical results in Fig.~\ref{fig:predHiggs}. For both curves, we take the ensemble average of 100 trajectories. The discrepancy between the the numerical and semi-analytical values can be ascribed to the approximations made in equations~\eqref{eq:rwa} and \eqref{eq:zeroModeApprox}. Moreover, we have ignored the $c$-axis dynamics and higher order terms as present in the Mexican hat potential, for example. Nonetheless, our estimate distils the effect of the fourth order coupling terms $\sim a^2 h^2$ on the Higgs frequency. For fixed model parameters, the Higgs frequency is shifted to a higher value because of thermally activated phase fluctuations in the $ab$-plane. In realistic systems, however, the chemical potential $\mu(T)$ decreases with temperature such that the Higgs frequency does not necessarily increase.

\bibliography{biblio}